\documentclass[11pt, draftcls, one column]{IEEEtran}

\usepackage{bbm}
\usepackage{graphicx}
\usepackage{subfigure}
\usepackage{color}
\usepackage{epsfig}
\usepackage{amssymb}
\usepackage{amsmath}
\usepackage{amsthm}
\usepackage{latexsym}
\usepackage{setspace}
\usepackage{bbm}
\usepackage{flushend}
\usepackage{dsfont}

\newcommand{\dP}{\mathrm{P}}
\newcommand{\dQ}{\mathrm{Q}}
\newcommand{\bQQ}[1]{{\mathrm{Q}}_{#1}}
\newcommand{\bQ}[2]{\mathrm{Q}_{#1}\left({#2}\right)}
\newcommand{\bP}[2]{\mathrm{P}_{#1}\left({#2}\right)}
\newcommand{\bPP}[1]{{\mathrm{P}}_{#1}}
\newcommand{\bPr}[1]{{\mathrm{P}}\left(#1\right)}

\newcommand{\tPP}[1]{{\tilde{\mathrm{P}}}_{#1}}
\newcommand{\tP}[2]{\tilde{\mathrm{P}}_{#1}\left({#2}\right)}

\newcommand{\mcf}{\mathrm{mcf}}
\newcommand{\mss}{\mathrm{mss}}
\renewcommand{\d}[2]{d\left( {#1},{#2}\right)}

\newcommand{\cA}{{\mathcal A}}

\newcommand{\cG}{{\mathcal G}}

\newcommand{\cK}{{\mathcal K}}

\newcommand{\cM}{{\mathcal M}}

\newcommand{\cT}{{\mathcal T}}

\newcommand{\cX}{{\mathcal X}}
\newcommand{\cY}{{\mathcal Y}}

\newcommand{\bF}{\mathbf{F}}

\newcommand{\bx}{\mathbf{x}}

\newcommand{\ep}{\epsilon}
\newcommand{\la}{\lambda}

\newcommand{\nn}{\nonumber}

\renewcommand{\atop}[2]{\genfrac{}{}{0pt}{1}{#1}{#2}}

\newtheorem{theorem}{Theorem}
\newtheorem{proposition}[theorem]{Proposition}
\newtheorem{corollary}[theorem]{Corollary}
\newtheorem*{corollary*}{Corollary}

\newtheorem{lemma}[theorem]{Lemma}
\newtheorem*{lemma*}{Lemmas}

\theoremstyle{remark}
\newtheorem{remark}{Remark}

\newtheorem{example}{Example}
\theoremstyle{definition}
\newtheorem{definition}{Definition}

\newcommand{\argmax}{\mathop{\rm argmax}\limits}

\newcommand{\indicator}{{\mathds{1}}}
 \IEEEoverridecommandlockouts
\usepackage{color}

\begin{document}
\title{Converses For Secret Key Agreement and Secure Computing}

\author{
\IEEEauthorblockN{Himanshu Tyagi$^\ast$} 
\and
\IEEEauthorblockN{Shun Watanabe$^\dag$} 
}

\maketitle

{\renewcommand{\thefootnote}{}\footnotetext{
\noindent$\ast$Department of Electrical Communication Engineering, Indian
Institute of Science, Bangalore 560012, India. 
Email: htyagi@ece.iisc.ernet.in

\noindent$^\dag$Department of Computer and Information Sciences, 
Tokyo University of Agriculture and Technology, Tokyo 184-8588, Japan. 
Email: shunwata@cc.tuat.ac.jp

Parts of this paper appeared in the proceedings of EUROCRYPT 2014 
and of IEEE International Symposium on Information Theory 2015.}
}

\begin{abstract}
We consider information theoretic secret key agreement and
secure function computation by multiple parties observing
correlated data, with access to an interactive public
communication channel. Our main result is an upper bound on
the secret key length, which is derived using a reduction of
binary hypothesis testing to multiparty secret key
agreement. Building on this basic result, we derive new
converses for multiparty secret key agreement. Furthermore,
we derive converse results for the oblivious transfer
problem and the bit commitment problem by relating them to
secret key agreement. Finally, we derive a necessary
condition for the feasibility of secure computation by
trusted parties that seek to compute a function of their
collective data, using an interactive public communication that
by itself does not give away the value of the function. In
many cases, we strengthen and improve upon previously known
converse bounds. Our results are single-shot and use only
the given joint distribution of the correlated observations.
 For the case when the correlated observations consist of
 independent and identically distributed (in time)
 sequences, we derive {\it strong} versions of previously
 known converses.
\end{abstract}
%
\section{Introduction}
Information theoretic cryptography relies on the
availability of correlated random observations to the
parties. Neither multiparty {\it secret key} (SK) agreement
nor secure computation is feasible if the observation of the
parties are mutually independent. In fact, SK agreement is
not feasible even when the observations are independent
across some partition of the set of parties\footnote{With
  restricted interpretations of {\it feasibility}, these
  observations appear across the vast literature on SK
  agreement and secure computation; see, for instance,
  \cite{Mau93}, \cite{AhlCsi93}, \cite{CsiNar04},
  \cite{RenWol05}, \cite{Kil00}, \cite{WinNasIma03},
  \cite{NasWin08}.}. As an extension of this principle, we
can expect that the efficiency of a cryptographic primitive
is related to ``how far'' the joint distribution of the
observations is from a distribution that renders the
observations independent (across some partition of the set
of parties).  We formalize this heuristic principle and
leverage it to bound the efficiency of using correlated
sources to implement SK agreement and secure computation. We
present {\it single-shot} converse results; in particular,
we do not assume that the observations of parties consist of
long sequences generated by an {\it independent and
  identically distributed} (IID) random
process\footnote{Throughout this paper, IID observations
  refer to observations that are IID {\it in time}; at each
  instant $t$, the observations of the parties are
  correlated.}.

In multiparty SK agreement, a set of parties observing
correlated {\it random variables} (RVs) seek to agree on
shared random bits that remain concealed from an
eavesdropper with access to a correlated side
information. The parties may communicate with each other
over a noiseless public channel, but the transmitted
communication will be available to the eavesdropper.  The
main tool for deriving our converse results is a reduction
argument that relates multiparty SK agreement to binary
hypothesis testing\footnote{This basic result was reported
  separately in \cite{TyaWat14}.}.  For an illustration of
our main idea, consider the two party case when the
eavesdropper observes only the communication between the
legitimate parties and does not observe any additional side
information. Clearly, if the observations of the legitimate
parties are independent, a SK cannot be generated. We upper
bound the length of SKs that can be generated in terms of
``how far" is the joint distribution of the observations of
the parties from a distribution that renders their
observations independent. Specifically, for this special
case, we show that the maximum length $S_\ep\left(X_1,
X_2\right)$ of a SK (for a given secrecy index $\ep$) is
bounded above as
\begin{eqnarray*}
S_\ep\left(X_1, X_2\right) \le -\log
\beta_{\ep+\eta}\big(\bPP{X_1 X_2}, \bPP{X_1} \times
\bPP{X_2}\big) + 2\log (1/\eta),
\end{eqnarray*}
where $\beta_{\ep}\big(\bPP{X_1 X_2}, \bPP{X_1} \times
\bPP{X_2}\big)$ is the optimal probability of error of type
II for testing the null hypothesis $\bPP{X_1 X_2}$ with the
alternative $\bPP{X_1} \times \bPP{X_2}$, given that the
probability of error of type I is smaller than $\ep$; this
$\beta_\ep$ serves as a proxy for ``distance'' between
$\bPP{X_1 X_2}$ and $\bPP{X_1} \times \bPP{X_2}$. Similarly,
in the general case of an arbitrary number of parties with
correlated side information at the eavesdropper, our main
result in Theorem
\ref{theorem:one-shot-converse-source-model} bounds the
secret key length in terms of the ``distance'' between the
joint distribution of the observations of the parties and
the eavesdropper and a distribution that renders the
observations of the parties conditionally independent across
some partition, when conditioned on the eavesdropper's side
information. This bound is a manifestation of the
aforementioned heuristic principle and is termed the {\it
  conditional independence testing} bound.

Our approach brings out a structural connection between SK
agreement and binary hypothesis testing\footnote{While the connection
between SK agreement and hypothesis testing established in this
paper is new, a similar connection between authentication and
hypothesis testing is natural to expect and is well-known
(see, for instance, \cite{Mau00})}. This is in the
spirit of \cite{PolPooVer10}, where a connection between
channel coding and binary hypothesis testing was used to
establish an upper bound on the rate of good channel codes
(see, also, \cite{WanRen12,HayNag03}). Also, our upper bound
is reminiscent of the {\it measure of entanglement} for a
quantum state proposed in \cite{VedPle98}, namely the
minimum distance between the density matrix of the state and
that of a disentangled state. This measure of entanglement
was shown to be an upper bound on the entanglement of
distillation in \cite{VedPle98}, where the latter is the
largest proportion of maximally entangled states that can be
distilled using a purification process
\cite{BenDiVSmoWot96}.

Using our basic result, we obtain new converses for SK
agreement, and also, for secure two-party computation by reducing SK
agreement to oblivious transfer and bit commitment. In many
cases, we strengthen and improve upon previously known
results. Our main contributions are summarized below.

\subsection{Secret key agreement}
For two parties, the problem of SK agreement from correlated
observations is well-studied. The problem was introduced by
Maurer \cite{Mau93} and Ahlswede and Csisz\'ar
\cite{AhlCsi93}, who considered the case where the parties
observe IID sequences. However, in certain applications it
is of interest to consider observations arising from a
single realization of correlated RVs. For instance, in
applications such as biometric and hardware authentication
(cf.~\cite{Pap01,DodOstReySmi08}), the correlated
observations consist of different versions of the biometric
and hardware signatures, respectively, recorded at the
registration and the authentication stages. To this end,
Renner and Wolf \cite{RenWol05} derived bounds on the length
of a SK that can be generated by two parties observing a
single realization of correlated RVs, using one-side
communication.

The problem of SK agreement with multiple parties, for the
IID setup, was introduced in \cite{CsiNar04} (also, see
\cite{CerMasSch02} for an early formulation).  In this work,
we consider the SK agreement problem for multiple parties
observing a single realization of correlated RVs. Our
conditional independence testing bound is a single-shot
upper bound on the length of SKs that can be generated by
multiple parties observing correlated data, using
interactive public communication\footnote{A single-shot
  upper bound using Fano's inequality for the length of a
  multiparty SK, obtained as a straightforward extension of
  \cite{CsiNar04, CsiNar08}, was reported in
  \cite{TyaWat13}.}. Unlike the single-shot upper bound in
\cite{RenWol05}, which is restricted to two parties with
one-way communication, we allow arbitrary interactive
communication between multiple parties. Asymptotically our
bound is tight -- its application to the IID case recovers
some previously known (tight) bounds on the asymptotic SK
rates. In fact, we strengthen the previously known
asymptotic results since we do not require the probability
of error in SK agreement or the secrecy index to be
asymptotically\footnote{Such bounds that do not require the
  probability of error to vanish to $0$ are called {\it
    strong converse} bounds \cite{CsiKor11}.}  $0$. See
Section \ref{section:SK-capacity} for a detailed discussion.
\subsection{Secure two-party computation}
The problem of secure two-party computation was introduced
by Yao in \cite{Yao82}.  Two (mutually untrusting) parties
seek to compute a function of their collective data, without
sharing anything more about their data than what is given
away by the function value. Several specific instances of
this general problem have been studied. We consider the
problems of {\it oblivious transfer} (OT) and {\it bit
  commitment} (BC), which constitute two basic primitives
for secure two-party computation.

OT between two parties is a mode of message transmission
``where the sender does not know whether the recipient
actually received the information" \cite{Rab81}. In this
paper, we consider the {\it one-of-two OT} problem
\cite{EveGolLem85} where the first party observes two
strings $K_0$ and $K_1$ of length $l$ each, and the second
party seeks the value of the $B$th string, $B\in
\{0,1\}$. The goal is to accomplish this task in such a
manner that $B$ and $K_{\overline{B}}$ remain concealed,
respectively, from Party 1 and Party 2. This simply stated
problem is at the heart of secure function computation as it
is well-known \cite{Kil88} that any secure function
computation task can be accomplished using the basic OT
protocol repeatedly (for recent results on the complexity of
secure function computation using OT, see
\cite{BeiIshKumKus14}). Unfortunately, information
theoretically secure OT is not feasible in the absence of
additional resources. On the bright side, if the parties
share a noisy communication channel or if they observe
correlated randomness, OT can be accomplished
(cf.~\cite{CreKil88, Cre97, AhlCsi13, NasWin08}). In this
paper, we consider the latter case where, as an additional
resource, the parties observe correlated RVs $X_1$ and
$X_2$.  Based on reduction arguments relating OT to SK
agreement, we derive upper bounds on the length $l$ of OT
that can be accomplished for given RVs $X_1, X_2$. The
resulting bound is, in general, tighter than that obtained
in \cite{WinWul12}. Furthermore, an application of our bound
to the case of IID observations shows that the upper bound
on the rate of OT length derived in \cite{NasWin08} and
\cite{AhlCsi13}\footnote{The asymptotic bound in
  \cite{AhlCsi13} is a special case of a more general
  asymptotic bound in \cite{RaoPra14}, which is based on
  tension technique introduced in \cite{PraPra10}. It is not
  clear if our approach can derive a single-shot version of
  the general bound in \cite{RaoPra14}.} is strong, $i.e.$,
the bound holds even without requiring asymptotically
perfect recovery.

We now turn to the BC problem, the first instance of which
was introduced by Blum in \cite{Blu83} as the problem of
flipping a coin over a telephone, when the parties do not
trust each other.  A bit commitment protocol has two
phases. In the first phase the committing party generates a
random bit string $K$, its ``coin flip''.  Subsequently, the
two parties communicate with each other, which ends the
first phase. In the second phase, the committing party
reveals $K$. A bit commitment protocol must forbid the
committing party from cheating and changing $K$ in the
second phase.  As in the case of OT, information
theoretically secure BC is not possible without additional
resources. We consider a version where two parties observing
correlated observations $X_1$ and $X_2$ want to implement
information theoretically secure BC using interactive public
communication. The goal is to maximize the length of the
committed string $K$.  By reducing SK agreement to BC, we
derive an upper bound on BC length which improves upon the
bound in \cite{RanTapWinWul11}. Furthermore, for the case of
IID observations, we derive a strong converse for BC
capacity; the latter is the maximum rate of BC length and
was characterized in \cite{WinNasIma03}.

\subsection{Secure computation with trusted parties}
In a different direction, we relate our result to the
following problem of {\it secure function computation with
  trusted parties} introduced in \cite{TyaNarGup11} (for an
early version of the problem, see \cite{OrlEl84}): Multiple
parties observing correlated data seek to compute a function
of their collective data. To this end, they communicate
interactively over a public communication channel, which is
assumed to be authenticated and error-free. It is required
that the value of the function be concealed from an
eavesdropper with access to the communication. When is such
a secure computation of a given function feasible? In
contrast to the traditional secure computation problem
discussed above, this setup is appropriate for applications
such as sensor networks where the legitimate parties are
trusted and are free to extract any information about each
other's data from the shared communication.  Using the
conditional independence testing bound, we derive a
necessary condition for the existence of a communication
protocol that allows the parties to reliably recover the
value of a given function, while keeping this value
concealed from an eavesdropper with access to (only) the
communication. In \cite{TyaNarGup11}, matching necessary and
sufficient conditions for secure computability of a given
function were derived for the case of IID observations. In
contrast, our necessary condition for secure computability
is single-shot and does not rely on the observations being
IID.
\subsection{Outline of paper}
The next section reviews some basic concepts that will be
used throughout this work.  The conditional independence
testing bound is derived in Section
\ref{section:conditional-independence-testing}.  In the
subsequent three sections, we present the implications of
this bound: Section \ref{section:SK-capacity} addresses
strong converses for SK capacity; Section
\ref{section:secure-computing} addresses converse results
for the OT and the bit commitment problem; and Section
\ref{section:secure-computing-trusted} contains converse
results for the secure computation problem with trusted
parties. The final section contains a brief discussion of
possible extensions.

\subsection{Notations}
For brevity, we use abbreviations SK, RV, and IID for secret
key, random variable, and independent and identically
distributed, respectively; a plural form will be indicated
by appending an `s' to the abbreviation. The RVs are denoted
by capital letters and the corresponding range sets are
denoted by calligraphic letters. The distribution of a RV
$U$ is given by $\bPP{U}$, when there is no confusion we
drop the subscript $U$. The set of all parties $\{1, ...,
m\}$ is denoted by $\cM$. For a collection of RVs $\{U_1,
.., U_m\}$ and a subset $A$ of $\cM$, $U_A$ denotes the RVs
$\{U_i,\, i \in A\}$. For a RV $U$, $U^n$ denotes $n$ IID
repetitions of the RV $U$. Similarly, $\bPP{}^n$ denotes the
distribution corresponding to the $n$ IID repetitions
generated from $\bPP{}$. All logarithms in this paper are to
the base $2$.

\section{Preliminaries}

\subsection{Secret keys}
Consider SK agreement using interactive public communication
by $m$ (trusted) parties. The $i$th party observes a
discrete RV $X_i$ taking values in a finite set $\cX_i$,
$1\leq i \leq m$.\footnote{The conditional independence
  testing bound given in Theorem
  \ref{theorem:one-shot-converse-source-model} remains valid
  even for continuous valued RVs. However, in general, the resulting bound
may not be achievable.} Upon making these
observations, the parties communicate interactively over a
public communication channel that is accessible by an
eavesdropper, who additionally observes a RV $Z$ such that
the RVs $\left(X_\cM, Z\right)$ have a distribution
$\bPP{X_\cM Z}$. We assume that the communication is
error-free and each party receives the communication from
every other party. Furthermore, we assume that the public
communication is authenticated and the eavesdropper cannot
tamper with it. Specifically, the communication is sent over
$r$ rounds of interaction. In the $j$th round of
communication, $1\leq j \leq r$, the $i$th party sends
$F_{ij}$, which is a function of its observation $X_i$, a
{\it locally generated} randomness\footnote{The RVs $U_1,
  ..., U_m$ are mutually independent and independent jointly
  of $(X_\cM, Z)$.} $U_i$ and the previously observed
communication
$$F_{11},..., F_{m1}, F_{12},..., F_{m2}, ..., F_{1j},...,
F_{(i-1)j}.$$ The overall interactive communication
$F_{11},...,F_{m1},...,F_{1r},...,F_{mr}$ is denoted by
$\bF$. Using their local observations and the interactive
communication $\bF$, the parties agree on a SK.

Formally, a SK is a collection of RVs $K_1, ..., K_m$, where
the $i$th party gets $K_i$, that agree with probability
close to $1$ and are concealed, in effect, from an
eavesdropper. Formally, the $i$th party computes a function
$K_i$ of $(U_i, X_i, \bF)$. Traditionally, the RVs $K_1,
..., K_m$ with a common range $\cK$ constitute an $(\ep,
\delta)$-SK if the following two conditions are satisfied
(for alternative definitions of secrecy, see
\cite{Mau93,Csi96,CsiNar04})
\begin{align}
\bPr{K_1 = \cdots = K_m} &\geq 1 - \ep,
\label{e:recoverability_old}
\\ \d{\bPP{K_1\bF Z}}{\bPP{\mathtt{unif}}\times \bPP{\bF Z}}
&\leq \delta,
\label{e:secrecy_old}
\end{align}
where $\bPP{\mathtt{unif}}$ is the uniform distribution on
$\cK$ and $\d{\bPP{}}{\bQQ{}}$ is the variational distance
between $\bPP{}$ and $\bQQ{}$ given by
$$\d{\bPP{}}{\bQQ{}} = \frac{1}{2}\sum_x|\bP{}{x} -
\bQ{}{x}|.$$ The first condition above represents the
reliable {\it recovery} of the SK and the second condition
guarantees {\it secrecy}. In this work, we use the
following alternative definition of a SK, which conveniently
combines the recoverability and the secrecy conditions
(cf.~\cite{Ren05}): The RVs $K_1, ..., K_m$ above constitute
an $\ep$-SK with common range $\cK$ if
\begin{align}
\d{\bPP{K_{\cal M}\bF
    Z}}{\bPP{\mathtt{unif}}^{({\cM})}\times \bPP{\bF Z}}
&\leq \ep,
\label{e:secrecy}
\end{align}
where
\begin{align*}
\bPP{\mathtt{unif}}^{({\cal M})}\left(k_{\cM}\right) =
\frac{\mathbbm{1}(k_1= \cdots = k_m)}{|\cK|}.
\end{align*}
In fact, the two definitions above are closely
related\footnote{Note that a SK agreement protocol that
  satisfies (\ref{e:secrecy}) {\it universally
    composable-emulates} an ideal SK agreement protocol (see
  \cite{Can01} for a definition).The emulation is with
  emulation slack $\ep$, for an environment of unbounded
  computational complexity.}.
\begin{proposition}\label{proposition:sec_equiv}
Given $0 \leq \ep, \delta < 1$, if $K_\cM$ constitute an
$(\ep, \delta)$-SK under (\ref{e:recoverability_old}) and
(\ref{e:secrecy_old}), then they constitute an
$(\ep+\delta)$-SK under (\ref{e:secrecy}).

Conversely, if $K_{\cM}$ constitute an $\ep$-SK under
(\ref{e:secrecy}), then they constitute an $(\ep, \ep)$-SK
under (\ref{e:recoverability_old}) and
(\ref{e:secrecy_old}).
\end{proposition}
Therefore, by the composition theorem in \cite{Can01}, the
complex cryptographic protocols using such SKs instead of
perfect SKs are secure.\footnote{A perfect SK refers to
  unbiased shared bits that are independent of
  eavesdropper's observations.}

We are interested in characterizing the maximum length
$\log|\cK|$ of an $\ep$-SK.
\begin{definition}\label{d:S_epsilon}
Given $0 \leq \ep < 1$, denote by $ S_\ep\left( X_\cM |
Z\right)$ the maximum length $\log|\cK|$ of an $\ep$-SK
$K_\cM$ with common range $\cK$.
\end{definition}

Our upper bound is based on relating the SK agreement
problem to a binary hypothesis testing problem; below we
review some basic concepts in hypothesis testing that will
be used.

\subsection{Hypothesis testing}\label{ss:hypothesis_testing}
Consider a binary hypothesis testing problem with null
hypothesis $\mathrm{P}$ and alternative hypothesis
$\mathrm{Q}$, where $\mathrm{P}$ and $\mathrm{Q}$ are
distributions on the same alphabet ${\cal X}$. Upon
observing a value $x\in \cX$, the observer needs to decide
if the value was generated by the distribution $\bPP{}$ or
the distribution $\mathrm{Q}$. To this end, the observer
applies a stochastic test $\mathrm{T}$, which is a
conditional distribution on $\{0,1\}$ given an observation
$x\in \cX$. When $x\in \cX$ is observed, the test
$\mathrm{T}$ chooses the null hypothesis with probability
$\mathrm{T}(0|x)$ and the alternative hypothesis with
probability $\mathrm{T}(1|x) = 1 - \mathrm{T}(0|x)$. For $0\leq \ep<1$, denote
by $\beta_\ep(\mathrm{P},\mathrm{Q})$ the infimum of the
probability of error of type II given that the probability
of error of type I is less than $\ep$, $i.e.$,
\begin{align}
\beta_\ep(\mathrm{P},\mathrm{Q}) := \inf_{\mathrm{T}\, :\,
  \mathrm{P}[\mathrm{T}] \ge 1 - \ep}
\mathrm{Q}[\mathrm{T}],
\label{e:beta-epsilon}
\end{align}
where
\begin{eqnarray*}
\mathrm{P}[\mathrm{T}] &=& \sum_x \mathrm{P}(x)
\mathrm{T}(0|x), \\ \mathrm{Q}[\mathrm{T}] &=& \sum_x
\mathrm{Q}(x) \mathrm{T}(0|x).
\end{eqnarray*}
We note two important properties of the quantity
$\beta_\ep(\mathrm{P},\mathrm{Q})$.
\begin{enumerate}
\item {\bf Data processing inequality.} Let $W$ be a
  stochastic mapping from $\cX$ to $\cY$, $i.e.$, for each
  $x\in \cX$, $W(\cdot | x)$ is a distribution on
  $\cY$. Then,
\begin{align} \label{e:dpi}
\beta_\ep(\mathrm{P},\mathrm{Q}) \le
\beta_\ep(\mathrm{P}\circ W, \mathrm{Q}\circ W),
\end{align}
where $(\mathrm{P}\circ W)(y) = \sum_x \bP{}{x}W(y| x)$.

\item {\bf Stein's Lemma.} (cf.~\cite[Theorem 3.3]{Kul68})
  For every $0< \ep <1$, we have
\begin{align}
\lim_{n\to\infty} - \frac{1}{n} \log
\beta_\ep(\mathrm{P}^n,\mathrm{Q}^n) = D(\mathrm{P} \|
\mathrm{Q}),
\label{e:Stein-lemma}
\end{align}
where $D(\mathrm{P}\|\mathrm{Q})$ is the Kullback-Leibler
divergence given by
\begin{eqnarray*}
D(\mathrm{P}\|\mathrm{Q}) = \sum_{x\in \cX} \mathrm{P}(x)
\log \frac{\mathrm{P}(x)}{\mathrm{Q}(x)},
\end{eqnarray*}
with the convention $0\log(0/0) = 0$.
\end{enumerate}

\subsection{Remarks on evaluation of $\beta_\ep(\dP, \dQ)$}
We close with a discussion on evaluating
$\beta_\ep(\mathrm{P},\mathrm{Q})$. Note that the expression
for $\beta_\ep(\mathrm{P},\mathrm{Q})$ in
\eqref{e:beta-epsilon} is a linear program, solving which
has a polynomial complexity in the size of the observation
space. A simple manipulation yields the following
computationally more tractable bound:
\begin{align}
-\log \beta_\ep(\mathrm{P},\mathrm{Q}) \leq \inf_{\gamma}
\gamma - \log\left(\bPP \gamma - \ep\right),
\label{eq:neyman-pearson-bound}
\end{align}
where
\[
\bPP \gamma = \bPr{\left\{x:
  \log\frac{\mathrm{P}(x)}{\mathrm{Q}(x)} \leq
  \gamma\right\}}
\]
When $\mathrm{P}$ and $\mathrm{Q}$ correspond to IID RVs,
the tail probability in \eqref{eq:neyman-pearson-bound} can
be numerically evaluated directly or can be approximated by
the B\'erry-Ess\'een theorem (cf.~\cite{Fel71}).  On the
other hand, numerical evaluation of the tail probability is
rather involved when $\mathrm{P}$ and $\mathrm{Q}$
correspond to Markov chains. For this case, a
computationally tractable and asymptotically tight bound on
$\beta_\ep(\mathrm{P}, \mathrm{Q})$ was established recently
in \cite{WatHay14}.  Also, by setting $\gamma =
D_\alpha(\mathrm{P}, \mathrm{Q}) +
\frac{1}{1-\alpha}\log(1-\ep -\ep^\prime)$, where $
D_\alpha(\mathrm{P}, \mathrm{Q}) $ is the R\'enyi's
divergence of order $\alpha>1$ and given by \cite{Ren61}
\begin{eqnarray*}
 D_\alpha(\mathrm{P}, \mathrm{Q}) = \frac{1}{\alpha
   -1}\log\sum_{x\in \cX}\mathrm{P}(x)^\alpha
 \mathrm{Q}(x)^{1-\alpha},
\end{eqnarray*}
the following simple bound on
$\beta_\ep(\mathrm{P}, \mathrm{Q})$ is obtained\footnote{For
  other connections between $\beta_\ep$ and R\'enyi's
  divergence, see \cite[Eqns. (3.37) and (3.38)]{Hayashi06}, 
\cite[Eqn. (29)]{PolVer10}.}:
\begin{align}
-\log \beta_\ep(\mathrm{P},\mathrm{Q}) \leq
D_\alpha(\mathrm{P}, \mathrm{Q}) +
\frac{1}{\alpha-1}\log\frac 1{1-\ep -\ep^\prime} + 
\log \frac 1{\ep^\prime}.
\label{eq:renyi-divergence-bound}
\end{align} 
A variant of this bound for the case of quantum observations 
was reported in \cite[Theorem 1]{OgawaN00}
(see, also, \cite[Eqn. (2.63)]{Hayashi06}). For the classical case,
the bound follows from the simple proof below: Denote by $A_\gamma$ the
set $\{x: \log\dP(x)/\dQ(x) \le \gamma\}$. Thus, for $\alpha>1$, 
\begin{align*}
1- \bPP \gamma &= \sum_{x\in A_\gamma^c} \dP(x)
\\
&= \sum_{x\in A_\gamma^c}  \dP(x)^\alpha \dP(x)^{1-\alpha }
\\
&< \sum_{x\in A_\gamma^c}  \dP(x)^\alpha
\dQ(x)^{1-\alpha }2^{(1-\alpha)\gamma}
\\
&\le 2^{(\alpha-1)D_\alpha(\dP, \dQ)+(1-\alpha)\gamma}
\\
&= (1-\ep -\ep^\prime),
\end{align*}
which further implies that $\bPP \gamma > \ep +\ep^\prime$. 
The bound \eqref{eq:renyi-divergence-bound} follows from 
\eqref{eq:neyman-pearson-bound}.
Note that while the bound \eqref{eq:renyi-divergence-bound} 
is not tight in general, as its corollary
we obtain Stein's lemma (see \eqref{e:Stein-lemma}).

Finally, we remark that when the condition
\begin{align} \label{eq:no-dispersion-condition}
\log\frac{\mathrm{P}(X)}{\mathrm{Q}(X)} = D(\mathrm{P} \|
\mathrm{Q})
\end{align} 
is satisfied with probability $1$ under $\mathrm{P}$, the
bound in \eqref{eq:neyman-pearson-bound} implies
\begin{align} \label{eq:no-dispersion-bound}
-\log \beta_\ep(\mathrm{P},\mathrm{Q}) \leq D(\mathrm{P} \|
\mathrm{Q}) + \log (1/(1-\ep)).
\end{align}

\subsection{Smooth min-entropy and smooth max-divergence}\label{ss:smoothing}
Given two RVs $X$ and $Y$, a central question of information
theoretic secrecy is (cf.~\cite{ImpLevLub89},
\cite{ImpZuc89}, \cite{BenBraCreMau95}): How many unbiased,
independent bits can be extracted from $X$ that are
unavailable to an observer of $Y$?  When the underlying
distribution is IID, the optimum rate of extracted bits can
be expressed in terms of Shannon entropies and is given by
$H(X|Y)$.  However, for our single-shot setup, {\it smooth
  min-entropy} introduced in \cite{RenWol05}, \cite{Ren05}
is a more relevant measure of randomness.  We use the
definition of smooth min entropy introduced\footnote{A
  review of the notion of smooth minimum entropy without the
  notations from quantum information theory can be also
  found in \cite{WatHay13}.}  in \cite{Ren05}; for a review
of other variations, see \cite{Tom12}.

We also review the {\it leftover hash lemma}
\cite{ImpLevLub89,BenBraCreMau95}, which brings out the
central role of {\it smooth min-entropy} in the answer to
the question above. Also, as a ``change of measure
companion'' for smooth min-entropy, we define {\it smooth
  max-divergence} and note that it satisfies the data
processing inequality.

\begin{definition}{\bf (min-entropy)}
The min-entropy of $\bPP{}$ is defined as
$$H_{\min}(\bPP{}) := \min_{x}\log\frac{1}{\bP{}{x}}.$$ For
distributions $\bPP{XY}$ and $\bQQ{Y}$, the conditional
min-entropy of $\bPP{XY}$ given $\bQQ{Y}$ is defined as
$$H_{\min}(\bPP{XY} | \bQQ{Y}) := \min_{x\,\in\, \cX,\,
  y\,\in\,
  \mathtt{supp}(\bQQ{Y})}\log\frac{\bQ{Y}{y}}{\bP{XY}{x,
    y}}.$$ Finally, the conditional minimum
entropy\footnote{There are other definitions of conditional
  min-entropy available in the literature. The form here is
  perhaps the most widely used and is appropriate for our
  purpose.}  of $\bPP{XY}$ given $Y$ is defined as
\begin{align}
H_{\min}(\bPP{XY} | Y) := \sup_{\bQQ{Y}}H_{\min}(\bPP{XY} |
\bQQ{Y}),
\label{e:min-entropy}
\end{align}
where the $\sup$ is over all $\bQQ{Y}$ such that
$\mathtt{supp}(\bPP{Y})\subseteq \mathtt{supp}(\bQQ{Y})$.
\end{definition}
Note that
\begin{align}
H_{\min}(\bPP{XY} | Y) &= - \inf_{\bQQ Y}\max_{x,y}\log
\frac{\bP{XY}{x,y}}{\bQ{Y}{y}} \nonumber \\ &= - \inf_{\bQQ
  Y}\max_{y}\log \frac{\bP{Y}{y}\max_x \bP {X|Y}
  {x|y}}{\bQ{Y}{y}} \nonumber \\ &= -\log \sum_{y}\bP
Y{y}\max_{x}\bP{X|Y}{x|y} - \inf_{\bQQ{Y}}\max_{y}\log
\frac{\tP Y y }{\bQ{Y}{y}} \nonumber \\ &= -\log \sum_{y}\bP
Y{y}\max_{x}\bP{X|Y}{x|y} - \inf_{\bQQ{Y}}D_{\max}(\tPP Y
\|\bQQ{Y}) \nonumber \\ &= -\log \sum_{y}\bP
Y{y}\max_{x}\bP{X|Y}{x|y}, \nonumber
\end{align}
where $\tP Y y
=\left(\sum_{y'}\bP{Y}{y'}\max_{x}\bP{X|Y}{x|y'}\right)^{-1}
\bP {Y}y\max_{x}\bP{X|Y}{x}$ and the final equality holds
since the max-divergence $D_{\max}(\dP\|\dQ)$ (see
Definition~\ref{d:max_divergence} below) is nonnegative and
equals $0$ if and only if $\dP = \dQ$. This alternative form
of conditional min-entropy was first derived in
\cite{KonigRS09} for a more general, quantum setup (see,
also, \cite[Theorem 2(ii)]{IwamotoS13}) and shows that
$H_{\min}(\bPP{XY} | Y)$ corresponds to the $-\log$ of the
{\it average conditional guessing probability} for $X$ given
$Y$. However, the original form in \eqref{e:min-entropy} is
more suited for our purpose.

The definition of min-entropy and conditional min-entropy
remain valid for all subnormalized, nonnegative functions
$\bPP{XY}$, $i.e.$, $\bPP{XY}$ such that
$$\sum_{x,y}\bP{XY}{x,y}\leq 1.$$ We need this extension and
the concept of smoothing, defined next, to derive tight
bounds.
\begin{definition}{\bf (Smooth min-entropy)} \label{definition:smooth-min-entropy}
Given $\ep\ge0$, the $\ep$-smooth conditional minimum
entropy of $\bPP{XY}$ given $Y$ is defined as
$$H_{\min}^\ep(\bPP{XY} | Y):= \sup_{\tPP{XY}:\,
  \d{\bPP{XY}}{\tPP{XY}}\,\leq\, \ep} H_{\min}(\tPP{XY} |
Y),$$ where the $\sup$ is over all subnormalized,
nonnegative functions $\tPP{XY}$. When $Y$ is a constant,
the $\ep$-smooth min-entropy is denoted by
$H_{\min}^\ep(\bPP{X})$.
\end{definition}
We now state the leftover hash lemma, which says that we can
extract $H_{\min}^\ep\left(\bPP{XY}| Y\right)$ unbiased,
independent bits from $X$ that are effectively concealed
from an observer of $Y$.
\begin{lemma}{\bf (Leftover hash) \cite{Ren05}}  \label{lemma:left-over-hash}
Given a joint distribution $\bPP{XY}$, for every $0\le 2\ep
< 1$ and $0< \eta$ there exists a mapping\footnote{A
  randomly chosen function from a $2$-universal hash family
  suffices.} $K: \cX\rightarrow \cK$ with $\log|\cK| =
\lfloor H_{\min}^\ep\left(\bPP{XY}| Y\right) -
2\log(1/2\eta)\rfloor$ such that
\begin{align}
\d{ \bPP{K(X)Y}}{\bPP{\mathtt{unif}}\times \bPP{Y}} \leq
2\ep + \eta.  \nonumber
\end{align}
\end{lemma}
Finally, we review smooth max-divergence, which was
introduced first in \cite{Dat09} for a quantum setting. The
method of smoothing in the following definition is slightly
different from the one in \cite{Dat09} and is tailored to
our purpose.
\begin{definition}{\bf (Smooth max-divergence)}\label{d:max_divergence}
The max-divergence between two distributions $\bPP{}$ and
$\bQQ{}$ is defined as
$$D_{\max}(\bPP{}\|\bQQ{}) :=
\max_{x}\,\log\frac{{\bP{}{x}}}{\bQ{}{x}},$$ with the
convention $\log (0/0) = 0$, and for $0< \ep < 1$, the
$\ep$-smooth max-divergence between $\bPP{}$ and $\bQQ{}$ is
defined as
\begin{align*}
D_{\max}^{\ep}(\bPP{}\|\bQQ{}) := \inf_{\atop{\tPP{}\,\leq\,
    \bPP{}:}{\tP{}{\cX} \,\geq\,
    1-\ep}}D_{\max}(\tPP{}\|\bQQ{}),
\end{align*} 
where the $\inf$ is over all subnormalized, nonnegative
functions $\tPP{}$ such that $\tP{}{x} \leq \bP{}{x}$ for
all $x\in\cX$ and $\sum_x\tP{}{x} \geq 1-\ep$.
\end{definition}
The following two properties of smooth max-divergence will
be used:
\begin{enumerate}
\item {\bf Data processing inequality.} For every stochastic
  mapping $W: \cX \rightarrow\cY$,
\begin{align}
D_{\max}^\ep(\bPP{}\circ W\| \bQQ{}\circ W ) \leq
D_{\max}^\ep(\bPP{}\| \bQQ{}).
\label{e:dpi_smooth_max_divergence}
\end{align}
Indeed, for every $\tPP{}$ such that $\tP{}{x} \leq
\bP{}{x}$ for all $x\in\cX$ and $\sum_x\tP{}{x} \geq 1-\ep$,
the following hold
\begin{align*}
(\tPP{} \circ W) (\cY) &\geq 1 - \ep,\\ (\tPP{} \circ W) (y)
  &\leq (\bPP{} \circ W) (y),\quad \forall\, y \in \cY.
\end{align*}
The property follows upon noting that for every $y \in \cY$
\begin{align*}
D_{\max}(\tPP{}\|\bQQ{}) &= \max_x \, \log
\frac{\tP{}{x}}{\bQ{}{x}} \\ &\geq \log \frac{(\tPP{} \circ
  W)(y)}{(\bQQ{} \circ W)(y)},
\end{align*}
since $\max_i (a_i/ b_i) \geq (\sum_i a_i / \sum_i b_i)$.
\item {\bf Convergence to Kullback-Leibler divergence.} For
  IID distributions $\bPP{}^n$ and $\bQQ{}^n$,
$$\lim_{n \rightarrow
    \infty}\frac{1}{n}\,D_{\max}^\ep(\bPP{}^n\| \bQQ{}^n) =
  D(\bPP{}\|\bQQ{}), \quad \forall\, 0< \ep <1.$$ The
  inequality `$\leq$' holds trivially if $D(\dP\|\dQ) =
  \infty$.  Thus, it suffices to prove it under the
  assumption that $D(\dP\|\dQ)$ is finite. To that end,
  consider $\tilde{\mathrm{P}}_n(\bx) =
  \mathrm{P}^n(\bx)\mathbbm{1}(\bx\in \cT_n)$, where $\cT_n$
  is the (strongly) typical set for $\bPP{}^n$
  (cf.~\cite{CsiKor11}). For a sequence $\bx \in \cX^n$ and
  an element $x\in \cX$, denote by $N(x|\bx)$ the number of
  occurrences of $x$ in $\bx$. Then, every sequence $\bx\in
  \cT_n$ satisfies (cf.~\cite{CsiKor11})
\begin{align}
\left|\frac {N(x|\bx)}n - \bP {}{x}\right| < \delta_n, \quad
x\in \cX,
\label{e:strong_typicality}
\end{align}
where $\delta_n \rightarrow 0$ as $n \rightarrow 0$ (for
precise conditions, see the {\it $\delta$-convention} in
\cite{CsiKor11}). Note that $\tPP n \leq \bPP{}^n$ and
$\tPP{n}(\cX^n) = \bPP{}^n(\cT_n) \geq 1 -\ep$ for all $n$
sufficiently large. Thus,
\begin{align}
\frac{1}{n}\,D_{\max}^\ep(\bPP{}^n\| \bQQ{}^n) &\leq
\frac{1}{n}\,D_{\max}(\tPP{n}\| \bQQ{}^n) 
\nn 
\\ &=
\max_{\bx \in \cT_n} \frac{1}{n}\log \frac{\bPP{}^n(\bx)}{
  \bQQ{}^n(\bx) } 
\nn 
\\ 
&= \max_{\bx \in
  \cT_n}\frac{1}{n}\sum_{i=1}^n \log \frac{\bPP{}(x_i)}{
  \bQQ{}(x_i) } 
\nn 
\\ 
&= \max_{\bx \in \cT_n}\sum_{x\in\cX} \frac{N(x|\bx)}{n}\log \frac{\bP {}{x}}{ \bQ{}{x} }
\nn 
\\ 
&\leq\sum_{x\in \cX}
\bP{}{x}\log \frac{\bP {}{x}}{ \bQ{}{x} } + o(1), \nn
\end{align}
where the last inequality follows from
\eqref{e:strong_typicality} under the assumption that
$D(\dP\|\dQ) < \infty$.

For the inequality in the other direction, suppose we are
given a $\tilde{\mathrm{P}}_n \leq \bPP{}^n$ with
$\tilde{\mathrm{P}}_n(\cX^n) \geq 1-\ep$. Then,
\begin{align}
\tilde{\mathrm{P}}_n(\cT_n) &= \tilde{\mathrm{P}}_n(\cX^n)-
\tilde{\mathrm{P}}_n(\cT_n^c) 
\nn 
\\ 
&\geq 1 - \ep -\tilde{\mathrm{P}}_n(\cT_n^c) 
\nn
\\
&\geq 1 - \ep -\bPP{}^n(\cT_n^c) 
\nn
\\
&\geq (1-\ep)/2,
\label{e:bound_on_tP}
\end{align}
for all $n$ sufficiently large. This further implies that there exists an
$\bx_0\in \cT_n$ such that
\[
\tPP{n}(\bx_0) \geq \bPP{}^n(\bx_0)(1-\ep)/2.
\]
Indeed, if not, then $\tPP{n}(\bx) < \bPP{}^n(\bx)(1-\ep)/2$
for all $\bx\in \cT_n$, which further implies
$\tPP{n}(\cT_n) < (1-\ep)/2$ contradicting
\eqref{e:bound_on_tP}.  Thus,
\begin{align*}
\frac 1
n\max_{\bx}\,\log\frac{\tilde{\mathrm{P}}_n(\bx)}{\mathrm{Q}^n(\bx)}
&\geq \frac 1
n\log\frac{\tilde{\mathrm{P}}_n(\bx_0)}{\mathrm{Q}^n(\bx_0)}
\\ &\geq \frac 1
n\log\frac{\bPP{}^n(\bx_0)}{\mathrm{Q}^n(\bx_0)} + \frac
1n\log \frac{1-\ep}2.
\end{align*}
For the case $D(\dP\|\dQ) =\infty$, there exists $x_+\in
\cX$ such that $\dP(x_+)>0$ and $\dQ(x_+) =0$. Since
$\bx_0\in \cT_n$, $N(x_+|\bx_0)>0$ and the right-side of the
inequality above, too, is infinity. On the other hand, if
$D(\dP\|\dQ)$ is finite, using \eqref{e:strong_typicality}
for the sequence $\bx_0\in \cT_n$, the right-side of the
inequality above is further bounded below by $D(\dP\|\dQ) -
o(1)$, which completes the proof.
\end{enumerate}

\section{The conditional independence testing bound}\label{section:conditional-independence-testing}
Converse results of this paper are based on an upper bound
on the maximum length $S_\ep\left(X_\cM | Z\right)$ of an
$\ep$-SK. We present this basic result here\footnote{The
  results of this section were presented in
  \cite{TyaWat14}.}.

Consider a (nontrivial) partition $\pi = \{\pi_1, ...,
\pi_l\}$ of the set $\cM$. Heuristically, if the underlying
distribution of the observations $\bPP{X_\cM Z}$ is such
that $X_\cM$ are conditionally independent across the
partition $\pi$ given $Z$, the length of a SK that can be
generated is $0$. Our approach is to bound the length of a
generated SK in terms of ``how far" is the distribution
$\bPP{X_\cM Z}$ from another distribution
$\mathrm{Q}^\pi_{X_{\cM}Z}$ that renders $X_\cM$
conditionally independent across the partition $\pi$ given
$Z$ -- the closeness of the two distributions is measured by
$\beta_{\ep}\big(\bPP{X_{\cM}Z},\mathrm{Q}^\pi_{X_{\cM}Z}\big)$.

Specifically, for a partition $\pi$ with $|\pi|\geq 2$
parts, let ${\cal Q}(\pi)$ be the set of all distributions
$\mathrm{Q}^\pi_{X_{\cM}Z}$ that factorize as follows:
\begin{align} \label{e:factorization}
\mathrm{Q}^\pi_{X_{\cM}| Z}(x_1,\ldots,x_m|z) =
\prod_{i=1}^{|\pi|}
\mathrm{Q}^\pi_{X_{\pi_i}|Z}(x_{\pi_i}|z).
\end{align}

\begin{theorem}[{{\bf Conditional independence testing bound}}] \label{theorem:one-shot-converse-source-model}
Given $0\leq \ep<1$, $0<\eta<1-\ep$, and a partition $\pi$
of $\cM$. It holds that
\begin{align}
S_\ep\left(X_\cM | Z\right) \le \frac{1}{|\pi|-1}
\bigg[-\log
  \beta_{\ep+\eta}\big(\bPP{X_{\cM}Z},\mathrm{Q}^\pi_{X_{\cM}Z}\big)
  + |\pi| \log(1/\eta)\bigg]
\label{eq:one-shot-converse-bound}
\end{align}
for all $\mathrm{Q}^\pi_{X_{\cM}Z} \in {\cal Q}(\pi)$.
\end{theorem} 
\begin{remark} 
Renner and Wolf \cite{RenWol05} derived a bound on the
length of a SK that can be generated by two parties using
one-way communication. A comparison of this bound with the
general bound in Theorem
\ref{theorem:one-shot-converse-source-model} is unavailable,
since the former involves auxiliary RVs and is difficult to
evaluate.
\end{remark}

\begin{remark}
For $m=2$ and $Z= \text{constant}$, the upper bound on the
length of a SK in Theorem
\ref{theorem:one-shot-converse-source-model} is related
closely to the {\it meta-converse} of Polyanskiy, Poor, and
Verd{\'u} \cite{PolPooVer10}. Indeed, a code for reliable
transmission of a message $M$ over a point-to-point channel
yields a SK for the sender and the receiver; the length of
this SK can be bounded by Theorem
\ref{theorem:one-shot-converse-source-model}. However, the
resulting bound is slightly weaker than the meta-converse
and does not yield the correct third order asymptotic term
(the coefficient of $\log n$) in the optimal size of
transmission codes \cite{TomTan13}.
\end{remark}

\begin{remark} \label{r:one-shot-converse}
The proof of Theorem
\ref{theorem:one-shot-converse-source-model} below remains
valid even when the secrecy condition \eqref{e:secrecy} is
replaced by the following more general condition:
$$\d{\bPP{K_\cM\bF Z}}{\bPP{\mathtt{unif}}^{(\cM)}\times
  \bQQ{\bF Z}} \leq \ep,$$ for {\it some} distribution
$\bQQ{\bF Z}$. In particular, upper bound
\eqref{eq:one-shot-converse-bound} holds even under the
relaxed secrecy criterion above.
\end{remark}

The key idea underlying our proof of
Theorem~\ref{theorem:one-shot-converse-source-model} is a
lower bound for $-\log \beta_\ep(\dP, \dQ)$ for a binary
hypothesis testing problem with observation space $\cK^m$,
null hypothesis $\dP$ given by
\begin{align}
\bPP{K_1K_2...K_m} = \bPP{\tt unif}^{(\cM)}
\label{e:null_multi}
\end{align}
and the alternative hypothesis $\dQ$ given by
\begin{align}
\bQQ{K_1K_2...K_m} = \prod_{i=1}^m \bQQ{K_i}
\label{e:alternative_multi}
\end{align}
namely the problem of testing if $K_1....K_m$ constitute a
perfectly correlated uniform randomness or are they mutually
independent.
\begin{lemma}\label{lemma:converse-key1}
For $\bPP{K_{\cM}}=\bPP{K_1...K_m}$ and $\bQQ{K_{\cM}} =
\bQQ{K_1...K_m}$ given in \eqref{e:null_multi} and
\eqref{e:alternative_multi}, it holds for every $0<\eta < 1$
that
\begin{align} 
\log |\cK| \le \frac{1}{m-1} \bigg[-\log
  \beta_{\eta}\big(\bPP{K_{\cM}},\mathrm{Q}_{K_{\cM}}\big) +
  m \log(1/\eta)\bigg].  \nn
\end{align}
\end{lemma}
\begin{proof} Consider the log-likelihood ratio test with threshold
  $\la$ given by
\[
\la = (m-1)\log|\cK| - m \log(1/\eta),
\]
$i.e.$, the deterministic test with the following acceptance
region (for the null hypothesis)
\begin{eqnarray*}
\cA := \left\{ k_{\cM} : \log
\frac{\bP{K_\cM}{k_\cM}}{\bQ{K_\cM}{k_\cM}} \ge \la\right\}.
\end{eqnarray*}
For this test, the probability of error of type II is
bounded above as
\begin{align}
\mathrm{Q}_{K_{\cM}}(\cA) &= \sum_{k_{\cM}\in \cA}
\bQ{K_\cM}{k_\cM} \nonumber \\ &\le 2^{-\lambda}
\sum_{k_{\cM}\in \cA} \bP{K_\cM}{k_\cM} \nonumber \\ &\le
2^{-\la} \nonumber \\ &= |\cK|^{1-m} \eta^{-m}.
 \label{e:bound-on-type-2}
\end{align}
On the other hand, the probability of error of type I is
given by
\begin{align} 
\bP{K_\cM}{\cA^c} &= \frac 1 {|\cK|} |\{k: {\bf k} = (k,
...,k)\in \cA^c\}| \nonumber\\ &= \frac{1}{|\cK|} \sum_k
\indicator\left({\bf k} \in \cA^c\right) \nonumber \\ &=
\frac{1}{|\cK|} \sum_k \indicator \left(
\bQ{K_{\cM}}{\mathbf{k}}|\cK|^m \eta^m > 1 \right),
\label{e:bound-on-type-1b}
\end{align}
where $\mathbf{k} := (k,\ldots,k)$ and the second equality
holds since $\cA^c$ consists of elements $k_\cM$ satisfying
\[
\frac{\bP{K_\cM}{k_\cM}}{\bQ{K_\cM}{k_\cM}} =
\frac{\indicator(k_1 = \cdots =k_m)}{|\cK|\bQ{K_\cM}{\cM}} <
2^{\la} = |\cK|^{m-1}\eta^m.
\]
The inner sum can be further upper bounded as
\begin{align}
\sum_k \indicator\left(\bQ{K_{\cM}}{\mathbf{k}}|\cK|^m
\eta^m > 1 \right) &\le \sum_k
\left(\bQ{K_{\cM}}{\mathbf{k}}|\cK|^m \eta^m \right)^{\frac
  1m} \nonumber \\ &= |\cK|\eta \sum_k
\bQ{K_{\cM}}{\mathbf{k}}^{\frac 1m} \nonumber \\ &=
|\cK|\eta \sum_k \prod_{i=1}^m\bQ{K_i}{k}^{\frac 1m}
\nonumber\\ &\le |\cK|\eta \prod_{i=1}^m\left(\sum_k
\bQ{K_i}{k}\right)^{\frac 1m}, \nonumber \\ &= |\cK|\eta.
\label{e:bound-on-type-1c}
\end{align}
where the second inequality above holds by H\"older's
inequality. Upon combining (\ref{e:bound-on-type-1b}) and
(\ref{e:bound-on-type-1c}) we obtain
\begin{eqnarray*}
\bP{K_\cM}{\cA^c} \leq \eta.
\end{eqnarray*}
Thus, we have a test with probability of error of type I
less than $\eta$ and the probability of error of type II
bounded as in (\ref{e:bound-on-type-2}). Therefore,
\[
\beta_{\eta}\big(\bPP{K_{\cM} },\bQQ{K_\cM}\big) \leq
|\cK|^{1-m} \eta^{-m},
\] 
which completes the proof.
\end{proof}
The distribution $\bPP{K_\cM}$ in \eqref{e:null_multi}
corresponds to a perfect secret key shared by $m$
parties. The next result extends
Lemma~\ref{lemma:converse-key1} to the case where not only
the key values $K_\cM$ but also the communication $\bF$ and
the eavesdropper's side information $Z$ are observed,
and the null hypothesis $\bPP{K_\cM\bF Z}$ corresponds
to an $\ep$-SK $K_{\cM}$.
\begin{lemma}\label{lemma:converse-key}
For an $\ep$-SK $K_\cM$ with a common range $\cK$ generated
using an interactive communication $\bF$, let $W_{K_\cM\bF |
  X_\cM Z}$ be the resulting conditional
distribution\footnote{The conditional distribution
  $W_{K_\cM\bF | X_\cM Z}$ is defined only for $(x_\cM, z)$
  with $\bP{X_\cM Z}{x_\cM,z}>0$.} on $\left(K_\cM,
\bF\right)$ given $\left(X_\cM,Z\right)$.  Then, for every
$0 < \eta < 1 -\ep$ and every $\bQQ{X_{\cM}Z} =
\prod_{i=1}^m\bQQ{X_i|Z}\bQQ{Z}$, we have
\begin{align} \label{e:converse-key}
\log |\cK| \le \frac{1}{m-1} \bigg[-\log
  \beta_{\ep+\eta}\big(\bPP{K_{\cM} \bF
    Z},\mathrm{Q}_{K_{\cM} \bF Z}\big) + m
  \log(1/\eta)\bigg],
\end{align}
where $\bPP{K_{\cM}\bF Z}$ is the marginal of
$\left({K_{\cM}, \bF, Z}\right)$ for the joint distribution
$$\bPP{K_{\cM}\bF X_\cM Z} = \bPP{X_{\cM}Z} W_{K_\cM\bF |
  X_\cM Z},$$ and $\mathrm{Q}^\pi_{K_{\cM}\bF Z}$ is the
corresponding marginal for the joint distribution
$$\mathrm{Q}_{K_{\cM}\bF X_\cM Z} = \mathrm{Q}_{X_{\cM}Z}
W_{K_\cM\bF| X_\cM Z}.$$
\end{lemma}
Also, we need the following basic property of interactive
communication from \cite{TyaNar13ii}, which will be used
throughout this paper (see, also, \cite[Lemma
  2.2]{AhlCsi93}, \cite[Lemma 17.18]{CsiKor11}).
\begin{lemma}[{{\bf Interactive communication property}}] \label{lemma:independence-preserve}
Given $\bQQ{X_{\cM}Z} = \prod_{i=1}^m\bQQ{X_i|Z}\bQQ{Z}$ and
an interactive communication $\bF$, the following holds:
\begin{eqnarray*}
\bQ{X_{\cM}|\bF Z}{x_{\cM} | f, z} = \prod_{i=1}^{m}
\bQ{X_i|\bF Z}{x_{i}|f, z},
\end{eqnarray*}
$i.e.$, conditionally independent observations remain so
when conditioned additionally on an interactive
communication. In particular, if $\mathrm{Q}_{X_1X_2| Z}
=\mathrm{Q}_{X_1|Z}\mathrm{Q}_{X_2| Z}$, then
\begin{align*}
\mathrm{Q}_{X_1X_2| \bF Z} = \mathrm{Q}_{X_1| \bF Z}
\times \mathrm{Q}_{X_2| \bF Z}.
\end{align*}
\end{lemma}
{\it Proof of Lemma \ref{lemma:converse-key}.}  The proof is
a simple modification of the proof of
Lemma~\ref{lemma:converse-key1}. First note that by
Lemma~\ref{lemma:independence-preserve} and the fact $K_i$
is a function of $(X_i, U_i)$ given $\bF$, we have
\[
\bQQ{K_\cM|\bF Z} = \prod_{i=1}^m\bQQ{K_i|\bF Z}.
\]
Thus, Lemma~\ref{lemma:converse-key1} applies with
distribution $\bQQ{K_\cM|\bF Z}$ in the role of $\dQ$ for
every $\bF = f, Z=z$, and consequently, for every $(f,z)$
there exists a set $\cA_{f,z}$ such that
\begin{align}
\bQ{K_\cM|\bF Z}{\cA_{f,z} | f, z} \leq
|\cK|^{1-m}\eta^{-m},
\label{e:type2_conditioned}
\end{align}
and
\begin{align}
\bPP{\tt unif}^{(\cM)}(\cA_{f,z}^c) \leq \eta.
\label{e:type1_conditioned}
\end{align}
We consider the following test for a binary hypothesis
testing problem with null hypothesis $\bPP{\tt unif}^{(\cM)}
\times\bPP{\bF Z}$ and alternative hypothesis $\bQQ{K_\cM\bF
  Z}$: For an observed $(k_\cM, f, z)$, we accept the null
hypothesis if $k_\cM\in \cA_{f,z}$ and alternative
otherwise. Using \eqref{e:type2_conditioned}, the
probability of error of type II is bounded above by
\[
\sum_{f,z}\bQ{\bF Z}{f, z}\bQ{K_\cM|\bF Z}{\cA_{f,z} | f,
  z} \leq |\cK|^{1-m}\eta^{-m},
\]
and by \eqref{e:type1_conditioned}, the probability of error
of type I is bounded above by
\[
\sum_{f,z}\bP{\bF Z}{f, z}\bPP{\tt unif}^{(\cM)}(\cA_{f,z})
\leq \eta.
\]
Finally, we consider the hypothesis testing problem with
null hypothesis $\bPP{K_\cM \bF Z}$ and alternative
hypothesis $\bQQ{K_\cM \bF Z}$ and apply the same test as
above. Clearly, the probability of error of type II remains
unchanged. Furthermore, in view of the secrecy condition
\eqref{e:secrecy}, the probability of error of type I will
increase by at most $\ep$, which completes the proof.  \qed
\vspace*{0.3cm}

{\it Proof of Theorem
  ~\ref{theorem:one-shot-converse-source-model}.}  We first
consider the partition $\pi$ with one element in each part,
$i.e.$, $\pi_i = \{i\}$ for $1\le i \le m$.  For this case,
it follows from Lemma~\ref{lemma:converse-key} and the data
processing inequality (\ref{e:dpi}) with $\mathrm{P} =
\bPP{X_{\cM}Z}$, $\mathrm{Q} =\mathrm{Q}^\pi_{X_{\cM}Z}$,
and $W = W_{K_\cM\bF | X_\cM Z}$, that for an $\ep$-SK
$K_\cM$ taking values in the set $\cK$
\begin{align}
\log |\cK| &\le \frac{1}{m-1} \bigg[-\log
  \beta_{\ep+\eta}\big(\bPP{K_{\cM}\bF
    Z},\mathrm{Q}^\pi_{K_{\cM}\bF Z}\big) + m
  \log(1/\eta)\bigg].  \nn \\ &\le \frac{1}{m-1} \bigg[-\log
  \beta_{\ep+\eta}\big(\bPP{X_{\cM}Z},\mathrm{Q}^\pi_{X_\cM
    Z}\big) + m \log(1/\eta)\bigg]
\label{e:SK_upper_bound_m}
\end{align}
for every $1< \eta < 1-\ep$.

To extend \eqref{e:SK_upper_bound_m} to an arbitrary
partition $\pi$, we claim that an $\ep$-SK for the original
model with $m$ parties yields an $\ep$-SK of the same length
for a model with $|\pi|$ parties with the $i$th party
observing $X_{\pi_i}$, $1\le i\le |\pi|$, and the
eavesdropper observing the RV $Z$ as before. The result
follows by applying the bound \eqref{e:SK_upper_bound_m} to
the new model with $|\pi|$ parties.

It only remains to prove the claim above. To that end, given
an $\ep$-SK $K_\cM$ for the original model, we define an
$\ep$-SK for the new model (with the $i$th party observing
$X_{\pi_i}$) as follows: The parties run the protocol for
generating $K_\cM$ with communication corresponding to any
party $j \in \pi_i$ in the original model transmitted by the
$i$th party in the new model. For each party $i$ in the new
model, we select a representative party $i_0 \in \pi_i$; for
concreteness, let $i_0$ be the smallest index in the set
$\pi_i$.  An $\ep$-SK for the new model is given by
$(K^\prime_1, ..., K^\prime_{|\pi|})$ where
 $K^\prime_i= K_{i_0}$ since, denoting by $\bPP{\tt
  unif}^{(\pi)}$ the distribution
\[
\bPP{\tt unif}^{(\pi)}(k_1,...,k_{|\pi|}) = \frac 1
    {|\cK|}\indicator(k_1 = \cdots =k_{|\pi|}),
\]
we have by the traingle inequality that
\begin{align*}
\d{\bPP{K^\prime_1...K^\prime_{|\pi|} \bF Z}}{\bPP{\tt
    unif}^{(\pi)} \times \bPP{\bF Z}} &\le \d{\bPP{K_1...K_m
    \bF Z}}{\bPP{\tt unif}^{(\cM)} \times \bPP{\bF Z}}
\\ &\le \ep,
\end{align*}
which completes the proof.  \qed
\section{Implications for secret key capacity}\label{section:SK-capacity}
For the SK agreement problem, a special case of interest is
when the observations consist of $n$ length IID sequences,
$i.e.$, the $i$th party observes $\left(X_{i1}, ...,
X_{in}\right)$ and the eavesdropper observes $\left(Z_1,
..., Z_n\right)$ such that the RVs $\{X_{\cM t},
Z_t\}_{t=1}^n$ are IID.  For this case, it is well known
that a SK of length proportional to $n$ can be generated;
the maximum rate $(\log|\cK_n|/n)$ of a SK is called the SK
capacity \cite{Mau93,AhlCsi93,CsiNar04}.

To present the results of this section at full strength, we
need to take recourse to the original definition of $(\ep,
\delta)$-SK given in \eqref{e:recoverability_old} and
\eqref{e:secrecy_old}.  In the manner of Definition
\ref{d:S_epsilon}, denote by $S_{\ep, \delta}(X_\cM|Z)$ the
maximum length of an $(\ep, \delta)$-SK. It follows from
Proposition \ref{proposition:sec_equiv} that $S_{\ep,
  \delta}(X_\cM|Z) \leq S_{\ep+\delta}(X_\cM|Z)$.
\begin{definition}{\bf (SK capacity)}
Given $0< \ep, \delta < 1$, the $(\ep,\delta)$-SK capacity
$C_{\ep,\delta}\left(X_\cM | Z\right)$ is defined by
\begin{eqnarray*}
C_{\ep,\delta}\left(X_\cM | Z\right) &:=&
\liminf_{n\to\infty} \frac{1}{n} S_{\ep,\delta}(X_\cM^n |
Z^n),
\end{eqnarray*}
where the RVs $\left\{X_{\cM t}, Z_t\right\}$ are IID for $1
\leq t \leq n$, with a common distribution $\bPP{X_\cM
  Z}$. The SK capacity $C\left(X_\cM | Z\right)$ is defined
as the limit
\begin{eqnarray*}
C\left(X_\cM | Z\right) := \lim_{\ep,\delta\rightarrow 0}
C_{\ep,\delta}\left(X_\cM | Z\right).
\end{eqnarray*}
\end{definition}

For the case when the eavesdropper does not observe any side
information, $i.e.$, $Z=constant$, the SK capacity for two
parties was characterized by Maurer \cite{Mau93} and
Ahlswede and Csisz{\'a}r \cite{AhlCsi93}. Later, the SK
capacity for a multiparty model, with $Z=$constant was
characterized by Csisz{\'a}r and Narayan
\cite{CsiNar04}. The general problem of characterizing the
SK capacity for arbitrary $Z$ remains open. Several upper
bounds for SK capacity are known
\cite{Mau93,AhlCsi93,MauWol99,RenWol03,CsiNar04,CsiNar08,GohAna10},
which are tight for special cases.

In this section, we derive a single-shot version of the
Gohari-Anantharam bound \cite{GohAna10} on the SK capacity
for two parties, which is the best known bound for this
case. Furthermore, for multiple parties, we establish a
strong converse for SK capacity, which shows that,
surprisingly, we cannot improve the rate of a SK by relaxing
the recoverability requirement \eqref{e:recoverability_old}
or the secrecy requirement \eqref{e:secrecy_old}.

\subsection{Converse results for two parties}
It was shown in \cite{GohAna10} that for two parties,
\begin{align}
C(X_1, X_2| Z) \leq \min_{U}I\left(X_1\wedge X_2| U\right) +
I(X_1, X_2\wedge U| Z).
\label{e:GA_bound}
\end{align}
The proof in \cite{GohAna10} relied critically on the
assumption that the RVs $\{\left(X_{\cM
  t},Z_t\right)\}_{t=1}^n$ are IID and does not apply to the
single-shot setup. The result below is a single-shot version
of \eqref{e:GA_bound} and is proved by relying only on the
structure of the SKs, without recourse to the {\it potential
  function approach}\footnote{In fact, a simple proof of
  \eqref{e:GA_bound} follows upon noting that for an optimum
  rate SK $(K_1,K_2)$ recoverable from a communication
  $\bF$, the SK capacity $C(X_1, X_2| Z)$ approximately
  equals $(1/n)H(K_1| \bF, Z^n) \leq (1/n)H(K_1|\bF,
  U^n,Z^n) + (1/n)I(K_1, \bF\wedge U^n|Z^n)$, which is
  further bounded above by $C(X_1, X_2| U) + I(X_1,X_2\wedge
  U| Z)$.}  of \cite{GohAna10}.

\begin{theorem}\label{theorem:single_shot_GA_bound}
For $0<\ep, \delta$ with $\ep+2\delta < 1$,
\begin{align*}
S_{\ep, \delta}(X_1,X_2|Z) \leq S_{\ep,
  2\delta+\eta}(X_1,X_2| Z, U) +
D_{\max}^{\xi}\left(\bPP{X_1X_2ZU}\|\bPP{X_1X_2Z}\bPP{U|
  Z}\right) + 2\log(1/2(\eta-\xi)) + 1,
\end{align*}
for every RV $U$ and every $0\leq \xi < \eta < 1-\ep
-2\delta$.
\end{theorem}
\noindent As corollaries, we obtain a single-shot version
and a strong version of the upper bound in
\eqref{e:GA_bound}, which does not require perfect
asymptotic recovery or perfect asymptotic secrecy.
\begin{corollary}[{{\bf Single-shot bound for SK length}}]\label{corollary:GA_singleshot} For $0<\ep, \delta$ with $\ep+2\delta < 1$,
\begin{align*}
S_{\ep, \delta}(X_1,X_2|Z) &\leq -\log\beta_{\ep+
  2\delta+\eta}(\bPP{X_1X_2ZU}, \bPP{X_1|ZU}\bPP{X_2ZU})
\\&\hspace*{0.5cm} +
D_{\max}^{\eta_1}\left(\bPP{X_1X_2ZU}\|\bPP{X_1X_2Z}\bPP{U|
  Z}\right)+ 4\log(1/(\eta-\eta_1 - \eta_2)) +1,
\end{align*}
for every RV $U$ and every $0\leq \eta_1+\eta_2 < \eta <
1-\ep -2\delta$.
\end{corollary}
\begin{corollary}[{{\bf Strong bound for SK capacity}}]\label{corollary:GA_strong} For $0\leq \ep, \delta$ with $\ep + 2\delta <1$,
\begin{align*}
C_{\ep, \delta}(X_1,X_2|Z) &\leq \min_{U} I\left(X_1\wedge
X_2| U\right) + I(X_1, X_2\wedge U| Z).
\end{align*}
\end{corollary}
We conclude this section with proofs. The core of Theorem
\ref{theorem:single_shot_GA_bound} is contained in the
following lemma.
\begin{lemma}\label{l:single_shot_GA_bound}
Let $(K_1, K_2)$ be an $(\ep, \delta)$-SK taking values in
$\cK$, recoverable from a communication $\bF$. Then,
\begin{align}
H_{\min}^{\delta+\xi/2}\left(\bPP{K_1\bF Z U} |\bF Z
U\right) \geq \log|\cK| - D_{\max}^{\xi}\left(\bPP{K_1\bF Z
  U}\|\bPP{K_1\bF Z}\bPP{U|Z}\right) \nonumber
\end{align}
for every RV $U$ and every $0\leq \xi < 1 - \ep- 2\delta$.
\end{lemma}
{\it Proof of Theorem \ref{theorem:single_shot_GA_bound}.}
Let $(K_1, K_2)$ be an $(\ep, \delta)$-SK taking values in
$\cK$. Then, by Lemma \ref{l:single_shot_GA_bound} and the
data processing property of smooth max-divergence
\eqref{e:dpi_smooth_max_divergence}, we get
\begin{align}
H_{\min}^{\delta+\xi/2}\left(\bPP{K_1\bF Z U} |\bF Z
U\right) \geq \log|\cK| - D_{\max}^{\xi}\left(\bPP{X_1X_2 Z
  U}\|\bPP{X_1X_2 Z}\bPP{U|Z}\right).  \nonumber
\end{align}
By the leftover hash lemma (see Section \ref{ss:smoothing}),
there exists a mapping $K^\prime$ of $\cK$ taking at least
$\log|\cK| - D_{\max}^{\xi}\left(\bPP{X_1X_2 Z
  U}\|\bPP{X_1X_2 Z}\bPP{U|Z}\right) - 2\log(1/2(\eta-\xi))
-1$ values and satisfying
$$\d{\bPP{K^\prime(K_1)\bF Z
    U}}{\bPP{\mathtt{unif}}\times\bPP{\bF Z U}} \leq 2\delta
+ \eta.$$ Therefore, $(K'(K_1),K'(K_2))$ constitutes an
$(\ep, 2\delta + \eta)$-SK for $X_1$ and $X_2$, when the
eavesdropper observes $(Z, U)$ and so,
$$\log|\cK| - D_{\max}^{\xi}\left(\bPP{X_1X_2 Z
  U}\|\bPP{X_1X_2 Z}\bPP{U|Z}\right) - 2\log(1/2(\eta-\xi))
- 1\leq S_{\ep, 2\delta+\eta}(X_1, X_2| Z, U).$$ \qed

Corollary \ref{corollary:GA_singleshot} follows by Theorem
\ref{theorem:one-shot-converse-source-model}.

{\it Proof of Corollary \ref{corollary:GA_strong}.} The
result follows by Corollary \ref{corollary:GA_singleshot}
upon using Stein's lemma (see Section
\ref{ss:hypothesis_testing}), along with the convergence
property of smooth max-divergence (see Section
\ref{ss:smoothing}).\qed

{\it Proof of Lemma \ref{l:single_shot_GA_bound}.}  By
definitions of $H_{\min}^{\delta+\xi/2}$ and
$D_{\max}^{\xi}$, it suffices to show that for every mapping
$T: (k_1,f,z,u) \mapsto [0,1]\allowbreak$ such that
\begin{align}
\sum_{k_1,f,z,u}\bP{}{k_1,f,z,u} T(k_1,f,z,u)\geq 1- \xi,
\label{e:T_definition}
\end{align}
here exist a subnormalized nonnegative function $\bQQ{K_1\bF
  ZU}$ and a distribution $\tilde{\mathrm{Q}}_{\bF Z U}$
satisfying the following:
\begin{align}
\d{\bPP{K_1\bF ZU}}{\bQQ{K_1\bF ZU}} &\leq \delta + \xi/2,
\label{e:smoothing}\\
H_{\min}\left(\bQQ{K_1\bF ZU}| \tilde{\mathrm{Q}}_{\bF Z
  U}\right) &= \log|\cK| - D_{\max}\left(\bPP{K_1\bF ZU}T
\|\bPP{K_1\bF Z}\bPP{U|Z}\right).
\label{e:minentropy_bound}
\end{align}
To that end, consider $\bQQ {K_1 \bF Z U}$ given by
\begin{align}
\bQ{} {k_1, f, z, u} := \bP{\mathtt{unif}}{k_1} \bP{}{f,z}
\bP{}{u|k_1, f,z} T(k_1,f,z,u),
\label{eq:definition-of-Q}
\end{align}
which is a valid subnormalized nonnegative function since
$T(k_1,f,z,u) \leq 1$. Furthermore, since
\[
\bP{} {k_1, f, z, u} =\bP{}{k_1, f,z} \bP{}{u|k_1, f,z},
\]
we get \eqref{e:smoothing} as follows:
\begin{align}
\d{\bPP{K_1\bF ZU}}{\bQQ{K_1\bF ZU}}
&\leq \d{\bPP{K_1\bF Z}}{\bPP{\tt unif}\bPP{\bF Z}} +
\sum_{k_1,f,z,u}\bP{}{k_1,f,z,u}(1 - T(k_1,f,z,u)) \nn
\\ &\leq \delta + \frac \xi 2, \nn
\end{align}
where the first inequality is by the triangle inequality and
the fact that $T(k_1, f,z,u)\leq 1$, and the last inequality
uses the secrecy condition \eqref{e:secrecy_old} and the
assumption \eqref{e:T_definition}.

Next, for $\tilde{\mathrm{Q}}_{\bF Z U}$ defined by
\begin{align}
\tilde{\mathrm{Q}}(f, z, u) := \bP{}{f,z}\bP{}{u|z}
\end{align}
and $Q_{K_1\bF Z U}$ defined in \eqref{eq:definition-of-Q},
observe that
\begin{align*}
\frac{\bQ{}{k_1, f, z, u}}{\tilde{\mathrm{Q}}(f, z, u)} &=
\bP{\mathtt{unif}}{k}\left[\frac{\bP{}{u| k_1, f,
      z}}{\bP{}{u|z}}\right]T(k_1, f, z, u) \\ &=
\bP{\mathtt{unif}}{k}\left[\frac{\bP{}{k_1, f, z,
      u}}{\bP{}{k_1, f,z}\bP{}{u|z}}\right]T(k_1, f, z, u)
\end{align*}
and so,
\begin{align*}
H_{\min}\left(\bQQ{K_1\bF Z U}| \tilde{\mathrm{Q}}_{\bF Z
  U}\right) &= \log|\cK| - \max_{k_1, f, z,
  u}\log\frac{\bP{}{k_1, f, z, u}T(k_1, f,z,u)}{\bP{}{k_1,
    f, z}\bP{}{u|z}},
\end{align*}
which is the same as \eqref{e:minentropy_bound}.  \qed

\subsection{Strong converse for multiple parties}

Now we move to the $m$ terminal case where the eavesdropper
gets no side information, $i.e.$, $Z = \mbox{constant}$.
With this simplification, the SK capacity
$C\left(X_\cM\right)$ for multiple parties was characterized
by Csisz\'ar and Narayan \cite{CsiNar04}. Furthermore, they
introduced the remarkable expression on the right-side of
(\ref{eq:divergence-expression}) below as an upper bound for
$C\left(X_\cM\right)$, and showed its tightness for $m=2,3$.
Later, the tightness of the upper bound for arbitrary $m$
was shown in \cite{ChaZhe10}; we summarize these
developments in the result below.
\begin{theorem}\label{theorem-capacity}\cite{CsiNar04,ChaZhe10} The SK capacity 
for the case when eavesdropper's side information $Z = \mbox{constant}$ is given by 
\begin{align}
C\left(X_\cM\right) = \min_{\pi} \frac{1}{|\pi| -1}
D\bigg(\mathrm{P}_{X_{\cM}} \bigg\| \prod_{i=1}^{|\pi|}
\mathrm{P}_{X_{\pi_i}} \bigg),
\label{eq:divergence-expression}
\end{align}
where the $\min$ is over all partitions $\pi$ of $\cM$.
\end{theorem}
This generalized the classic result of Maurer \cite{Mau93}
and Ahlswede and Csisz\'ar \cite{AhlCsi93}, which
established that for two parties, $C\left(X_1,X_2\right) =
D\left(\bPP{X_1X_2} \middle\| \bPP{X_1} \times
\bPP{X_2}\right) = I\left(X_1\wedge X_2\right)$.

The converse part of Theorem \ref{theorem-capacity} relied
critically on the fact that $\ep_n+\delta_n \rightarrow 0$ as $n \rightarrow \infty$. Below we strengthen the
converse and show that the upper bound for SK rates implied
by Theorem \ref{theorem-capacity} holds even when
$(\ep_n,\delta_n)$ is fixed. Specifically, for $0<
\ep,\delta$ with $\ep+\delta<1$ and $Z=\mbox{constant}$, an
application of Theorem
\ref{theorem:one-shot-converse-source-model} to the IID RVs
$X_\cM^n$, with $\mathrm{Q}^\pi_{X_\cM^n} =
\prod_{i=1}^{|\pi|} \bPP{X_{\pi_i}}^n$, yields
\begin{eqnarray*}
S_{\ep,\delta}\left(X_1^n, ..., X_m^n\right)\leq
\frac{1}{|\pi|-1} \left[-\log
  \beta_{\ep+\delta+\eta}\left(\bPP{X_{\cM}}^n,\prod_{i=1}^{|\pi|}
  \bPP{X_{\pi_i}}^n\right) + |\pi| \log(1/\eta)\right],
\end{eqnarray*}
 where $\eta < 1- \ep -\delta$. Therefore, using Stein's
 Lemma (see (\ref{e:Stein-lemma})) we get
\begin{eqnarray*}
C_{\ep,\delta}\left(X_\cM\right) &\leq& \frac{1}{|\pi|-1}
\liminf_{n \rightarrow \infty}-\frac{1}{n}\log
\beta_{\ep+\delta+\eta}\left(\bPP{X_{\cM}}^n,\prod_{i=1}^{|\pi|}
\bPP{X_{\pi_i}}^n\right) \\ &=& \frac{1}{|\pi| -1}
D\bigg(\mathrm{P}_{X_{\cM}} \bigg\| \prod_{i=1}^{|\pi|}
\mathrm{P}_{X_{\pi_i}} \bigg).
\end{eqnarray*}

Also, note that if $\ep+\delta\ge1$, the SK rate can be
infinity.  Indeed, consider a $(0,1)$-SK where party $1$
generates a RV $K_1$ uniformly over a set $\cK$ and sends it
to the other parties over the public communication channel,
and a $(1,0)$-SK where the $i$th party generates $K_i$
uniformly over $\cK$ using its local randomness $U_i$
(without any public communication).  If $\ep+\delta \ge 1$,
the SK which equals the $(0,1)$-SK above with probability
$(1-\ep)$ and the $(1,0)$-SK above with probability $\ep$
constitutes an $(\ep, 1-\ep)$-SK of length $\log|\cK|$, and
therefore, also an $(\ep, \delta)$-SK of the same
length. Since $\cK$ was arbitrary, the length of the
resulting $(\ep,\delta)$-SK can be arbitrarily large.

Thus, we have established the following {\it strong
  converse} for the SK capacity when $Z=\mbox{constant}$.
\begin{corollary}[{{\bf Strong converse for SK capacity}}] \label{corollary-strong-converse-for-multi}
Given $0< \ep, \delta <1$, the $(\ep,\delta)$-SK capacity
$C_{\ep,\delta}\left(X_\cM\right)$ is given by
\begin{eqnarray*}
C_{\ep,\delta}\left(X_\cM\right) = \min_{\pi} \frac{1}{|\pi|
  -1} D\bigg(\mathrm{P}_{X_{\cM}} \bigg\|
\prod_{i=1}^{|\pi|} \mathrm{P}_{X_{\pi_i}} \bigg), \quad
\text{ if } \ep+\delta <1,
\end{eqnarray*}
and
\begin{eqnarray*}
C_{\ep,\delta}\left(X_\cM\right) = \infty, \quad \text{ if }
\ep+\delta \geq 1.
\end{eqnarray*}
\end{corollary}

\section{Implications for secure two-party computation} \label{section:secure-computing}
\begin{figure}[t]
\centering
\setlength{\unitlength}{.4mm}
\begin{picture}(200, 130)
\put(30,120){\textcolor{blue}{Conditional Independence Testing}}
\put(110,100){reduction in Theorem \ref{theorem:one-shot-converse-source-model}}
\thicklines
\put(100,115){\vector(0,-1){30}}
\put(61,75){\textcolor{blue}{Secret Key Agreement}}
\thicklines
\put(80,70){\vector(-1,-1){30}}
\thicklines
\put(120,70){\vector(1,-1){30}}
\put(-10,60){Lemmas \ref{l:OT_reduction1} and \ref{l:OT_reduction2}}
\put(135,60){Lemma \ref{l:reduction_SK_BC}}
\put(20,30){\textcolor{red}{Oblivious Transfer}}
\put(110,30){\textcolor{red}{Bit Commitment}}
\put(17,15){\dashbox{2}(162,35)[s]{}}
\put(36,5){\textcolor{red}{Secure Two-Party Computation}}

  \end{picture}
  \caption{Depiction of our reduction arguments. }
  \label{fig:reduction}
\end{figure}
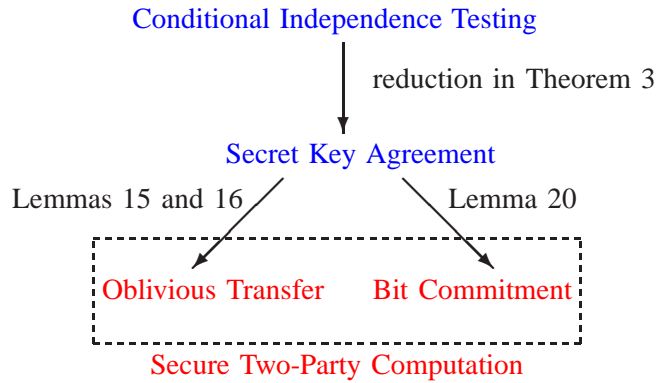

In this section, we consider secure computation by
two (mutually untrusting) parties.  First introduced by Yao
in \cite{Yao82}, these problems have propelled the research
in cryptography over the last three decades. In particular,
we will consider the {\it oblivious transfer} and the {\it
  bit commitment} problem, the two basic primitives for
secure two-party computation. We will look at the information theoretic
versions of these problems where, as an additional resource,
the parties observe correlated RVs $X_1$ and $X_2$. Our
converse results are based on reduction arguments which
relate these problems to the SK agreement problem, enabling
the application of Theorem
\ref{theorem:one-shot-converse-source-model} (see
Fig.~\ref{fig:reduction}).

\subsection{Maximum common function and minimum sufficient statistic}
To state our results, we need the notions of {\it maximum
  common function} and {\it minimum sufficient statistic}.
The notion of maximum common function was introduced in
\cite{GacKor73} as a measure of ``common information'' of
random variables $X_1$ and $X_2$. Its role in secrecy was
first highlighted in \cite{WolWul08} (see, also,
\cite{TyaNarGup11, NTW15} for different roles of the maximum
common function in secrecy and privacy.)  Operationally, the
maximum common function of $X_1$ and $X_2$ is defined as
follows.
\begin{definition}[{{\bf Maximum Common Function}}]
A {\it common function} of $X_1$ and $X_2$ is a random
variable $U$ for which there there exist functions
$\phi_1(X_1)$ and $\phi_2(X_2)$ such that
$\bPr{U=\phi_1(X_1) = \phi_2(X_2)} = 1$. The maximum common
function\footnote{By definition, it is unique up to
  relabeling.} of $X_1$ and $X_2$, denoted by
$\mcf(X_1,X_2)$, is a common function of $X_1$ and of $X_2$
such that every common function $U$ of $X_1$ and $X_2$
is a function of $\mcf(X_1,X_2)$, $i.e.$, $H(U|\mcf(X_1,
X_2))=0$.
\end{definition}
In fact, \cite{GacKor73} characterized $\mcf(X_1, X_2)$ and
showed that it corresponds to the following equivalence relation
on $\cX_1$ (or a similarly defined equivalence relation on
$\cX_2$)
\begin{align*}
x_1 \sim x_1^\prime \Leftrightarrow & \,\exists \,
x_{21},...,x_{2k} \in \cX_2 \text{ and } x_{12}, ...,
x_{1k}\in \cX_1~\text{s.t. with }x_{11}= x_1 \text{ and }
x_{1(k+1)}=x_1^\prime, \\ &
\bP{X_1X_2}{x_{1i},x_{2i}}\bP{X_1X_2}{x_{1(i+1)},x_{2i}} > 0
\text{ for }1\le i \le k.
\end{align*}

The role of minimum sufficient statistic in secrecy was
highlighted in \cite{WolWul08} as well. We give its
operational definition below.
\begin{definition}[{{\bf Minimum Sufficient Satistics}}]
A {\em sufficient statistic} for $X_2$ given $X_1$ is a random variable
$U$ such that there exists a function $U= g(X_1)$ such that 
the Markov chain $X_1 \text{---} U \text{---} X_2$ holds.
The minimum sufficient statistics for $X_2$ given $X_1$,
denoted by $\mss(X_2|X_1)$, is a sufficient statistics
for $X_2$ given $X_1$ such that it is a function of every sufficient statistic $U$
for $X_2$ given $X_1$, i.e., $H(\mss(X_2|X_1)|U)  =0$.
\end{definition}
An exact characterization of $\mss(X_2| X_1)$ is available,
too, and it corresponds to the following equivalence
relation on $\cX_1$ (cf. \cite{FitWolWul04},
\cite{KamAna10}, \cite{Tya13}):
\begin{align}
x_1\sim x_1' \Leftrightarrow \bP{X_2|X_1}{x_2|x_1} =
\bP{X_2|X_1}{x_2|x_1'}, \quad \text{for all } x_2\in \cX_2.
\nn
\end{align}

\subsection{Oblivious transfer}	

We present bounds on the efficiency of implementing
information theoretically secure one-of-two OT using
correlated randomness.  Suppose that party 1 generates $K_0$
and $K_1$, distributed uniformly over $\{0,1\}^l$, and party
2 generates $B$, distributed uniformly over $\{0,1\}$, as
inputs to an OT protocol. The RVs $K_0, K_1,$ and $B$ are
assumed to be
mutually independent\footnote{Strictly speaking, OT refers
  to the problem where the strings $K_0, K_1$ and the bit
  $B$ are fixed. The randomized version here is sometimes
  referred as {\it oblivious key transfer} (see \cite{Bea95,
    WolWul06}) or {\it fully randomized oblivious transfer}
  (see \cite{WehCurSchLo10, KonigWW12}), and they are
  equivalent to OT.}.  The goal of an OT protocol is for
Party 2 to obtain $K_B$ in such a manner that $B$ is
concealed from Party 1 and $K_{\overline{B}}$ is concealed
from party 2, where $\overline{B} = 1 \oplus B$.
Furthermore, Party $i$ observes the RV $X_i$, $i=1,2$, as a
resource to implement an OT protocol, where RVs $(X_1, X_2)$
are independent jointly of $(K_0, K_1, B)$.  During the
protocol, the parties are allowed to communicate
interactively.  In general, the parties are allowed to use
local randomization; for simplicity of presentation, we
restrict ourselves to protocols without local
randomization. However, as pointed-out in Remark
\ref{r:OT_reduction2} below, our results remain valid even
when local randomization is allowed.
\begin{definition}{(\bf Oblivious transfer)} 
A protocol realizing an $(\ep, \delta_1,
\delta_2)$-OT (for a passive adversary\footnote{Here,
  ``passive adversary'' refers to an ``honest but curious''
  adversary that follows the protocol, but is curious to
  know the other party's input.  Since we consider only
  converse results for OT, this assumption only strengthens
  our results and they remain valid for more powerful,
  active adversaries.})  of length $l$ consists of an interactive 
communication $\bF$ and an estimate $\hat{K} = \hat{K}(X_2,
B, \bF)$ by Party 2 such that the following conditions are satisfied:
\begin{align}
\bPr{K_B \neq \hat{K}} &\leq \ep,
\label{e:OT_recovery}
\\ \d{\bPP{K_{\overline{B}}X_2B\bF}}{\bPP{K_{\overline{B}}}\times\bPP{X_2B\bF}}
&\leq \delta_1,
\label{e:OT_secrecy1}
\\ \d{\bPP{B K_0 K_1 X_1\bF}}{\bPP{B}\times\bPP{K_0 K_1
    X_1\bF}} &\leq \delta_2,
\label{e:OT_secrecy2}
\end{align}
where $\overline{B} = 1\oplus B$. The first condition above
denotes the reliability of OT, while the second and the
third conditions ensure secrecy for party 1 and 2,
respectively.  Denote by $L_{\ep,\delta_1, \delta_2}(X_1,
X_2)$ the largest $l$ such that a protocol realizing
an $(\ep, \delta_1, \delta_2)$-OT of length $l$ exists.
\end{definition}
When the underlying observations $X_1, X_2$ consist of
$n$-length IID sequences $X_1^n, X_2^n$ with common
distribution $\bPP{X_1X_2}$, it is known that
$L_{\ep,\delta_1, \delta_2}(X_1^n, X_2^n)$ may grow linearly
with $n$ (cf.~\cite{NasWin08, AhlCsi13}); the largest rate
of growth is called the OT capacity.
\begin{definition}[{{\bf OT capacity}}]  For $0< \ep < 1$, the
 $\ep$-OT capacity of $(X_1,X_2)$ is defined\footnote{For
    brevity, we use the same notation for SK capacity and OT
    capacity; the meaning will be clear from the
    context. Similarly, the notation $L$, used here to
    denote the optimal OT length, is also used to denote the
    optimal BC length in the next section.}  as
\begin{align*}
C_\epsilon(X_1, X_2) = \lim_{\delta_1,\delta_2\to0}
\liminf_{n \to \infty} \frac{1}{n}\,L_{\ep,\delta_1,
  \delta_1}(X_1^n, X_2^n).
\end{align*}
Then, the OT capacity is defined as
\begin{align*}
C(X_1,X_2) = \lim_{\ep \to 0} C_\epsilon(X_1, X_2).
\end{align*}
\end{definition}
The main result of this section is an upper bound on
$L_{\ep,\delta_1, \delta_2}(X_1, X_2)$. Consequently, we
recover the upper bound on $C(X_1,X_2)$ due to Ahlswede and
Csisz\'ar derived in \cite{AhlCsi13}. In fact, we show that
the upper bound is ``strong" and applies to $C_\ep(X_1,X_2)$
for every $0< \ep < 1$.

Heuristically, OT is feasible
only when the observations of the two parties are correlated.
However, no party should have an advantage over the other,
and only that portion of correlated randomness observed 
by a party is useful which cannot be determined by the other party.
Drawing on this heuristic, two different bounds for OT length are possible,
each based on relating OT length to ``how far'' the joint distribution $\bPP{X_1X_2}$  
of the observed correlated randomness is from a useless distribution.
The choice of the useless distribution is different in both bounds. In the first 
bound,
we consider a distribution such that $X_1$ and $X_2$ are independent given
$V_0 = \mcf(X_1, X_2)$. For such distributions, once the shared knowledge
of each party is factored out, no correlation is available to facilitate OT.
In the second bound, we consider distributions where $V_1 = \mss(X_2|X_1)$
can be determined by $X_2$. Note that for such distributions 
the factorization $\bPP{V_1V_1X_2} = \bPP{V_1|X_2}\bPP{V_1|X_2}\bPP{X_2}$ holds.
Such distributions are useless for OT since the essential
part of $X_1$ that is correlated with $X_2$, namely $V_1$,
can be determined by $X_2$, thereby giving an advantage to
Party 2. As in the case of SK agreement, we shall measure the
distance between two distributions using $\beta_\ep$. In
fact, our proof entails reducing SK agreement to OT; the
reductions used for the two bounds are different.

\begin{theorem}[{{\bf Single-shot bound for OT length}}]\label{theorem:OT_upper_bounds}
For RVs $X_1, X_2$, $V_0 = \mcf(X_1, X_2)$ and $V_1 =
\mss(X_2| X_1)$, the following inequalities hold:
\begin{align}
L_{\ep,\delta_1, \delta_2}(X_1, X_2) &\leq
-\log\beta_{\eta}\left(\bPP{X_1X_2V_0},\bPP{X_1|
  V_0}\bPP{X_2| V_0}\bPP{V_0}\right) + 2\log(1/\xi),
\label{e:OT_bound1}
\\ L_{\ep,\delta_1, \delta_2}(X_1, X_2) &\leq
-\log\beta_{\eta}\left(\bPP{V_1V_1X_2},\bPP{V_1|
  X_2}\bPP{V_1| X_2}\bPP{X_2}\right) + 2\log(1/\xi),
\label{e:OT_bound2}
\end{align}
for all $\xi>0$ with $\eta = \ep + \delta_1 +2\delta_2 +\xi
< 1$.
\end{theorem}
\begin{corollary}[{{\bf Strong bound for OT capacity}}]\label{c:strong_bound_OT_capacity}
For $0< \ep < 1$, the $\ep$-OT capacity of $(X_1, X_2)$
satisfies
\begin{align} \label{eq:strong-bound-OT-capacity}
C_\ep(X_1, X_2) \leq \min\{ I(X_1\wedge X_2| V_0),\,H(V_1|
X_2)\},
\end{align}
where $V_0 = \mcf(X_1, X_2)$ and $V_1 = \mss(X_2|X_1)$.
\end{corollary}
The proof of Theorem \ref{theorem:OT_upper_bounds} entails
reducing two SK agreement problems to OT\footnote{A
  reduction of SK to OT in a computational secrecy setup
  appeared in \cite{GerKanMalReiVis00}.}. The bound
\eqref{e:OT_bound1} is obtained by recovering $K_B$ as a SK,
while \eqref{e:OT_bound2} is obtained by recovering
$K_{\overline{B}}$ as a SK; we note these two reductions as
separate lemmas below.
\begin{lemma}[{{\bf Reduction 1 of SK agreement to OT}}]\label{l:OT_reduction1}
Consider SK agreement for two parties observing $X_1$ and
$X_2$, respectively, with the eavesdropper observing $V_0 =
\mcf(X_1, X_2)$. Given a protocol realizing an $(\ep,
\delta_1, \delta_2)$-OT of length $l$, there exists a
protocol for generating an $(\ep + \delta_1 + 2\delta_2)$-SK
of length $l$. In particular,
$$L_{\ep,\delta_1, \delta_2}(X_1, X_2) \leq S_{\ep+\delta_1
  + 2\delta_2}(X_1, X_2| V_0).$$
\end{lemma}
\begin{lemma}[{{\bf Reduction 2 of SK agreement to OT}}]\label{l:OT_reduction2}
Consider two party SK agreement where Party 1 observes
$X_1$, Party 2 observes $(V_1, X_2) = (\mss(X_2|X_1), X_2)$
and the eavesdropper observes $X_2$. Given a protocol
realizing an $(\ep, \delta_1, \delta_2)$-OT of length $l$,
there exists a protocol for generating an
$(\ep+\delta_1+2\delta_2)$-SK of length $l$. In particular,
$$L_{\ep,\delta_1, \delta_2}(X_1, X_2) \leq S_{\ep+\delta_1
  + 2\delta_2}(X_1, (V_1, X_2)| X_2).$$
\end{lemma}
\begin{remark} 
Underlying the proof of $C(X_1, X_2)\leq I(X_1\wedge X_2)$
in \cite{AhlCsi13} was a reduction of SK agreement to OT,
which is extended in our proof below to prove
\eqref{e:OT_bound1}.  In contrast, the proof of the bound
$C(X_1, X_2)\leq H(X_1| X_2)$ in \cite{AhlCsi13} relied on
manipulations of entropy terms. Below we give an alternative
reduction argument to prove \eqref{e:OT_bound2}.
\end{remark}

\begin{remark}
In general, our bounds are stronger than those presented in
\cite{WinWul12}. For instance, the latter is loose when the
observations consist of mixtures of IID RVs.  Further, while
both \eqref{e:OT_bound2} and \cite[Theorem 5]{WinWul12}
(specialized to OT) suffice to obtain the second bound in
Corollary \ref{c:strong_bound_OT_capacity}, in contrast to
\eqref{e:OT_bound1}, \cite[Theorem 2]{WinWul12} does not
yield the first bound in Corollary
\ref{c:strong_bound_OT_capacity}.
\end{remark}

\begin{remark} \label{r:OT_reduction2}
For simplicity of presentation, we did not allow local
randomization in the formulation above. However, it can be
easily included as a part of $X_1$ and $X_2$ by replacing
$X_i$ with $(X_i, U_i)$, $i=1,2$, where $U_1, U_2, (X_1,
X_2)$ are mutually independent. Since our proofs are based
on reduction of SK agreement to OT, by noting that
$\mss(X_2, U_2|X_1, U_1) = \mss(X_2| X_1)$ and that the
availability of local randomness does not change our upper
bound on SK length in Theorem
\ref{theorem:one-shot-converse-source-model}, the results
above remain valid even when local randomness is available.
\end{remark}

\begin{remark}
The $(\ep, \delta_1, \delta_2)$-OT capacity $C_{\ep, \delta_1, \delta_2}(X_1, X_2)$
can be defined,
without requiring $\delta_1, \delta_2$ to go to $0$ as in
the definition of $C_\ep(X_1, X_2)$. However, the right-side
of \eqref{eq:strong-bound-OT-capacity} constitutes an upper
bound for $C_{\ep, \delta_1, \delta_2}(X_1, X_2)$
only when
$\ep+\delta_1+2\delta_2 <1$, and establishing the validity 
of this bound for\footnote{For $\ep+\delta_1+\delta_2 \ge 1$, 
$C_{\ep, \delta_1, \delta_2}(X_1, X_2)$ can be shown to be unbounded in 
the manner of the discussion preceding
  Corollary~\ref{corollary-strong-converse-for-multi}.}
$\ep+\delta_1+2\delta_2 \ge 1$ remains an open problem.
\end{remark}

We prove Lemmas \ref{l:OT_reduction1} and
\ref{l:OT_reduction2} next. The proof of Theorem
\ref{theorem:OT_upper_bounds} follows by Theorem
\ref{theorem:one-shot-converse-source-model}, along with the
Markov relation $X_1 \text{---}V_1\text{---}X_2$ and the
data processing inequality \eqref{e:dpi}; the corollary
follows by Stein's Lemma (see Section
\ref{ss:hypothesis_testing}).

{\it Proof. of Lemma \ref{l:OT_reduction1}.}  Let $\hat{K}$
be the estimate of $K_B$ formed by Party 2. The following
protocol generates an $(\ep + \delta_1 + 2\delta_2)$-SK of
length $l$
\begin{enumerate}
\item[(i)] Party 1 generates two random strings $K_0$ and
  $K_1$ of length $l$, and Party 2 generates a random bit
  $B$. Two parties run the OT protocol, and Party 2 obtains
  an estimate $\hat{K}$ of $K_B$.

\item[(ii)] Party 2 sends $B$ over the public channel.

\item[(iii)] Using $B$, Party 1 computes $K_B$.
\end{enumerate}
We will show that the RVs $K_B, \hat{K}$ constitute an
$(\ep+\delta_1+2\delta_2)$-SK. The reliability for this
SK is guaranteed since both parties agree on
$K_B$ with probability greater than $1-\ep$. For establishing secrecy,
note that if Party 2 sends $\overline{B}$ instead of $B$,
the eavesdropper cannot determine $K_B$ from
$(\overline{B},\bF)$ by the secrecy condition for Party 1.  
On the other hand, by the secrecy condition for
Party 2, the overall observation $(K_0,K_1,X_1,\bF)$ of Party 1
has roughly the same distribution even when 
$B$ is replaced by $\overline{B}$. Thus, the eavesdropper cannot
determine $K_B$ from $(B,\bF)$ as well.

Formally, by Proposition \ref{proposition:sec_equiv} and
Remark \ref{r:one-shot-converse}, it suffices to show that
for some distribution $Q_{V_0\bF B}$ (see Remark
\ref{r:one-shot-converse}),
\begin{align*}
\d{\bPP{K_BV_0\bF B}}{\bPP{\mathtt{unif}}\times \bQQ{V_0\bF
    B}} \leq \delta_1+2\delta_2.
\end{align*}  
Observe that condition \eqref{e:OT_secrecy2} is the same as
\begin{align}
\d{\bPP{K_0K_1X_1\bF| B = 0}}{\bPP{K_0K_1X_1\bF| B =
    1}} \leq 2\delta_2.
\label{e:OT_secrecy2'}
\end{align}
Let $\bQ{V_0\bF B}{v,f, b} = \bP{V_0\bF | B}{v,
  f|\overline{b}}\bP{B}{b}$. Then,
\begin{align*}
&\d{\bPP{K_BV_0\bF B}}{\bPP{\mathtt{unif}}\times \bQQ{V_0\bF
      B}} \\ &= \frac{1}{2}\sum_b \d{\bPP{K_bV_0\bF|
      B=b}}{\bPP{\mathtt{unif}}\times \bQQ{V_0\bF|B=b}}
  \\ &= \frac{1}{2}\sum_b \d{\bPP{K_bV_0\bF|
      B=b}}{\bPP{\mathtt{unif}}\times
    \bPP{V_0\bF|B=\overline{b}}} \\ &\leq
  \frac{1}{2}\sum_b\left[ \d{\bPP{K_bV_0\bF|
        B=\overline{b}}}{\bPP{\mathtt{unif}}\times
      \bPP{V_0\bF|B=\overline{b}}} + \d{\bPP{K_bV_0\bF|
        B=b}}{\bPP{K_bV_0\bF| B=\overline{b}}}\right] \\ &=
  \d{\bPP{K_{\overline{B}}V_0\bF
      B}}{\bPP{\mathtt{unif}}\times \bPP{V_0\bF B}} +
  \frac{1}{2}\sum_b \d{\bPP{K_bV_0\bF| B=b}}{\bPP{K_bV_0\bF|
      B=\overline{b}}} \\ &\leq \delta_1+ 2\delta_2,
\end{align*}
where the last inequality uses \eqref{e:OT_secrecy1} and
\eqref{e:OT_secrecy2'}, together with the fact that $V_0$
is a function of $X_2$ as well as $X_1$. \qed

{\it Proof. of Lemma \ref{l:OT_reduction2}.}  The following
protocol generates an $(\ep+\delta_1+2\delta_2)$-SK of
length $l$.
\begin{enumerate}
\item[(i)] Party 1 generates two random strings $K_0$ and
  $K_1$ of length $l$, and Party 2 generates a random bit
  $B$. Two parties run the OT protocol.

\item[(ii)] Upon observing $\bF$, Party 2 samples
  $\tilde{X}_2$ according to the distribution\\ $\bP{X_2|
  V_1 B \bF}{\cdot | V_1, \overline{B}, \bF}$.

\item[(iii)] Party 2 sends $B$ over the public channel.

\item[(iv)] Party 1 computes $K_{\overline{B}}$ and Party 2
  computes $\tilde{K} = \hat{K}(\tilde{X}_2, \overline{B},
  \bF)$.
\end{enumerate}
We will show that the RVs $K_{\overline{B}}, \tilde{K}$
constitute an $(\ep+\delta_1+2\delta_2)$-SK.  Heuristically,
this protocol entails Party 2 emulating $\tilde{X}_2$,
pretending that the protocol was executed for $\overline{B}$
instead of $B$. Since the communication of Party 1 is
oblivious of the value of $B$, plugging $\tilde{X}_2$ into
$\hat{K}$ will lead to an estimate of $K_{\overline{B}}$
provided that the emulated $\tilde{X}_2$ preserves the joint
distribution.

By Proposition \ref{proposition:sec_equiv} and
\eqref{e:OT_secrecy1}, it suffices to show that
\begin{align}
\bPr{K_{\overline{B}}\neq \tilde{K}} \leq \ep + 2\delta_2.
\label{e:OT_proof}
\end{align}
To that end, note
\begin{align*}
&\bPr{K_{\overline{B}}\neq \tilde{K}}
  \\ &=\frac{1}{2}\sum_{k,\,b,\, v, f} \bP{K_{\overline{b}}
    V_1\bF| B}{k, v, f| b} \bPr{\hat{K}(X_2, \overline{b},
    f) \neq k\mid V_1=v, B = \overline{b}, \bF =f} \\ &\leq
  \frac{1}{2}\sum_{k,\,b,\, v, f} \bP{K_{\overline{b}}
    V_1\bF| B}{k, v, f| \overline{b}} \bPr{\hat{K}(X_2,
    \overline{b}, f) \neq k\mid V_1=v, B = \overline{b}, \bF
    =f} + 2\delta_2 \\ &= \frac{1}{2}\sum_{k,\,b,\, v, f}
  \bP{K_{b} V_1\bF| B}{k, v, f| b} \bPr{\hat{K}(X_2, b, f)
    \neq k\mid V_1=v, B = b, \bF =f} + 2\delta_2
  \\ &=\bPr{K_B \neq \hat{K}} + 2\delta_2.
\end{align*}
where the inequality uses \eqref{e:OT_secrecy2'} and the
last equality uses the Markov relation $X_2
\text{---}V_1B\bF\text{---} K_0K_1$, which holds in the view
of the interactive communication property of Lemma
\ref{lemma:independence-preserve}; \eqref{e:OT_proof}
follows by \eqref{e:OT_recovery}.  \qed

\subsection{Bit commitment}
Two parties observing correlated observations $X_1$ and
$X_2$ want to implement information theoretically secure BC
using interactive public communication, $i.e.$, the first
party seeks to report to the second the results of a series
of coin tosses that it conducted at its end in such a manner
that, at a later stage, Party 2 can detect if Party 1 was
lying \cite{Blu83}. Formally, a BC protocol consists of two
phases: the {\it commit phase} and the {\it reveal
  phase}. In the commit phase, Party 1 generates a random
string $K$, distributed uniformly over $\{0,1\}^l$ and
independent jointly of $(X_1, X_2)$. Furthermore, the two
parties communicate interactively with each other using an
interactive communication $\bF$. In the
reveal phase, Party 1 ``reveals'' its data, $i.e.$, it sends
$X_1'$ and $K'$, claiming these were its initial choices of
$X_1$ and $K$, respectively. Subsequently, Party 2 applies a
(randomized) test function $T = T(K', X_1', X_2, \bF)$,
where $T=0$ and $T=1$, respectively, indicate $K'=K$ and
$K'\neq K$.
\begin{definition}[{{\bf Bit commitment}}] A protocol 
realizing an $(\ep, \delta_1, \delta_2)$-BC of length $l$ 
consists of an interactive communication $\bF$ to be sent during the
commit phase and a $\{0,1\}$-valued randomized test
function $T$ to be used in the reveal phase such that the following
conditions are satisfied:
\begin{align}
\bPr{T(K, X_1, X_2, \bF) \neq 0} &\leq \ep,
\label{e:BC_recovery}
\\ \d{\bPP{KX_2\bF}}{\bPP{K}\times\bPP{X_2\bF}} &\leq
\delta_1,
\label{e:BC_secrecy1}
\\ \bPr{T(K', X_1', X_2, \bF) = 0, K'\neq K} &\leq \delta_2,
\label{e:BC_secrecy2}
\end{align}
for any choice of RVs $K^\prime$ and $X_1^\prime$ that have
the same range-sets as
$K$ and $X_1$, respectively\footnote{ Note that
this restriction is valid since a dishonest Party 1 seeks
to replace $K$ with $K^\prime$ in the reveal phase, 
without being caught by Party 2.}, and satisfy
\begin{align*}
(K^\prime,X_1^\prime) \text{---} (K,X_1,\bF) \text{---} X_2.
\end{align*}
The first condition above is the {\it soundness condition},
which captures the reliability of BC when Party 1 is
honest. The next condition is the {\it hiding condition},
which ensures that Party 2 cannot ascertain the secret in
the commit phase. Finally, the {\it binding condition} in
\eqref{e:BC_secrecy2} restricts the probability with which
Party 1 can cheat in the reveal phase. 
Denote by
$L_{\ep,\delta_1, \delta_2}(X_1, X_2)$ the largest 
$l$ such that a protocol realizing an $(\ep, \delta_1,
\delta_2)$-BC of length $l$ exists.
\end{definition}
For $n$-length IID sequences $X_1^n, X_2^n$ generated from
$\bPP{X_1X_2}$, the largest rate of $L_{\ep,\delta_1,
  \delta_2}(X_1, X_2)$ is called the BC capacity.
\begin{definition}[{{\bf BC capacity}}] For $0< \ep, \delta_1, \delta_2<1$, the $(\ep, \delta_1, \delta_2)$-BC capacity of $(X_1, X_2)$ is defined as 
\begin{align*}
C_{\ep, \delta_1, \delta_2}(X_1, X_2) = \liminf_{n \to
  \infty} \frac{1}{n}L_{\ep, \delta_1, \delta_2}(X_1^n,
X_2^n).
\end{align*}
The BC capacity is defined as
\begin{align*}
C(X_1, X_2) = \lim_{\ep, \delta_1, \delta_2 \rightarrow
  0}C_{\ep, \delta_1, \delta_2}(X_1, X_2).
\end{align*}
\end{definition}
The following result of Winter, Nascimento, and Imai
\cite{WinNasIma03} (see, also, \cite[Chapter
  8]{TuySkoKev07}) gives a simple formula for $C(X_1, X_2)$.
\begin{theorem}\cite{WinNasIma03} For RVs $X_1, X_2$, let $V_1 = \mss(X_2|X_1)$. The BC capacity is 
given by
\begin{align*}
C(X_1, X_2) = H(V_1 | X_2).
\end{align*}
\end{theorem}
In this section, we present an upper bound on $L_{\ep,
  \delta_1, \delta_2}(X_1, X_2)$, which in turn leads to a
strong converse for BC capacity.
\begin{theorem}[{{\bf Single-shot bound for BC length}}]\label{theorem:BC_single_shot_bound}
Given $0< \ep, \delta_1, \delta_2$,
$\ep+\delta_1+\delta_2<1$, for RVs $X_1, X_2$ and $V_1 =
\mss(X_2| X_1)$, the following inequality holds:
\begin{align*}
L_{\ep, \delta_1, \delta_2}(X_1, X_2) \leq
-\log\beta_{\eta}\left(\bPP{V_1V_1X_2},\bPP{V_1|
  X_2}\bPP{V_1| X_2}\bPP{X_2}\right) + 2\log(1/\xi),
\end{align*}
for all $\xi$ with $\eta = \ep+\delta_1+\delta_2+\xi$.
\end{theorem}
\begin{corollary}[{{\bf Strong converse for BC capacity}}]
\label{c:BC_strong_converse} For $0< \ep, \delta_1, \delta_2$, $\ep+\delta_1+\delta_2<1$, the $(\ep, \delta_1, \delta_2)$-BC capacity satisfies
\begin{align*}
C_{\ep, \delta_1, \delta_2}(X_1, X_2) \leq H(V_1| X_2),
\end{align*}
where $V_1 = \mss(X_2| X_1)$.
\end{corollary}
Theorem \ref{theorem:BC_single_shot_bound} is obtained by a
reduction of SK agreement to BC, which is along the lines of
\cite{WinNasIma03,ImaMorNasWin06,RanTapWinWul11}; the
following lemma captures the resulting bound.
\begin{lemma}[{{\bf Reduction of SK to BC}}]\label{l:reduction_SK_BC}
For $0< \ep, \delta_1, \delta_2$, $\ep+\delta_1+\delta_2<1$,
it holds that
\begin{align*}
L_{\ep, \delta_1, \delta_2}(X_1, X_2) \leq
S_{\ep+\delta_1+\delta_2}(X_1, (V_1, X_2) | X_2),
\end{align*}
where $V_1 = \mss(X_2| X_1)$.
\end{lemma}
\begin{remark}\label{r:BC_strong_converse}
While local randomization was not allowed in the foregoing
discussion, as before (see Remark \ref{r:OT_reduction2}) our
results do not change with the availability of local
randomness.
\end{remark}

\begin{remark}
For $\ep, \delta_1, \delta_2>0$, $\ep+\delta_1+\delta_2<1$,
the following bound on $L_{\ep, \delta_1, \delta_2}(X_1,
X_2)$ was derived in \cite[Lemma 4]{RanTapWinWul11}:
\begin{align*}
L_{\ep, \delta_1, \delta_2}(X_1, X_2) \leq \frac{H(V_1|X_2)
  + h(\delta_1) +
  h(\epsilon+\delta_2)}{1-\epsilon-\delta_1-\delta_2},
\end{align*}
where $h(\cdot)$ is the binary entropy function.  However,
this bound is weaker than Theorem
\ref{theorem:BC_single_shot_bound}, in general, and is not
sufficient for deriving Corollary
\ref{c:BC_strong_converse}.
\end{remark}
Theorem \ref{theorem:BC_single_shot_bound} follows by using
Lemma \ref{l:reduction_SK_BC} with Theorem
\ref{theorem:one-shot-converse-source-model}, along with the
Markov relation $X_1 \text{---}V_1\text{---}X_2$ and the
data processing inequality \eqref{e:dpi}; the Corollary
\ref{c:BC_strong_converse} follows by Stein's Lemma (see
Section \ref{ss:hypothesis_testing}).  We prove Lemma
\ref{l:reduction_SK_BC} below.

{\it Proof of Lemma \ref{l:reduction_SK_BC}.} The reduction
argument presented here is along the lines of
\cite[Proposition 9]{ImaMorNasWin06} (see, also, \cite[Lemma
  4]{RanTapWinWul11}).  Given an
$(\epsilon,\delta_1,\delta_2)$-BC of length $l$, consider SK
agreement by two parties observing $X_1$ and $(V_1, X_2)$,
respectively, with the eavesdropper observing $X_2$. To
generate a SK, the parties run the commit phase of the BC
protocol, $i.e.$, Party 1 generates $K\sim
\mathtt{unif}\{0,1\}^l$ and the parties send the interactive
communication $\bF$. We show that the committed secret $K$
constitutes a $(\ep+\delta_2,\delta_1)$-SK. Indeed, by the
hiding condition \eqref{e:BC_secrecy1}, the SK $K$
satisfies the secrecy condition \eqref{e:secrecy_old} with
$\delta = \delta_1$. To establish the reliability of this SK,
we show that, roughly, $K$ is the unique string
which is compatible with $(V_1,X_2,\bF)$, namely that any
other string will fail the test $T$, since otherwise 
a dishonest Party 1 can change the
string in the reveal phase, contradicting the
binding condition. Thus, Party 2 can obtain an estimate of
$K$ by finding the unique string that is compatible with
$(V_1,X_2,\bF)$.

Formally, we complete the proof by showing that there exists $\hat{K}
= \hat{K}(V_1, X_2, \bF)$ such that
\begin{align}
\bPr{\hat{K} \neq K} \leq \ep+ \delta_2.
\label{e:BC_to_show}
\end{align}
 To that end, let $(\hat{k}, \hat{x}_1) = (\hat{k}(v,f), \hat{x}_1(v,f))$ be a
 function of $(v,f)$ given by
 \begin{align*}
 (\hat{k}, \hat{x}_1) &= \argmax_{k, x_1}
   \bPr{T(k, x_1, X_2, \bF) = 0 \mid V_1=v, \bF = f} \\ &=
   \argmax_{k, x_1}\sum_{x_2}\bP{X_2|V_1\bF}{x_2| v, f}
   \bPr{T(k, x_1, x_2,f) = 0},
 \end{align*}
 and let $(\hat{K}, \hat{X}_1) = (\hat{k}(V_1,\bF),
 \hat{x}_1(V_1,\bF))$.  Note that while the estimated secret
 $\hat{K}$ is a function of $(v,f)$ and
does not depend on $X_2$ directly, the latter is needed to
 facilitate the communication $\bF$ in the emulation of the
 commit phase. For $ (\hat{K}, \hat{X}_1)$ as above, we get
 \begin{align*}
&\bPr{T(\hat{K}, \hat{X}_1, X_2,\bF) = 0} \\ &=\sum_{v, f}
   \bP{V_1\bF}{v, f}\sum_{x_2}\bP{X_2|V_1\bF}{x_2| v, f}
   \bPr{T(\hat{k}(v,f), \hat{x}_1(v,f), x_2, f) = 0}
   \\ &\geq \sum_{v, f} \bP{V_1\bF}{v, f}\sum_{k,x_1}\bP{K,
     X_1|V_1\bF}{k, x_1| v, f}
   \sum_{x_2}\bP{X_2|V_1\bF}{x_2| v, f} \bPr{T(k, x_1, x_2,
     f) = 0} \\ &= \bPr{T({K}, {X}_1, X_2,\bF) = 0} \\ &\geq
   1- \ep,
 \end{align*}
 where the first inequality uses the definition of
 $(\hat{k}(v,f), \hat{x}_1(v,f))$ and the second equality
 uses the Markov relation $KX_1 \text{---} V_1\bF \text{---}
 X_2$, which holds in the view of the interactive
 communication property of Lemma
 \ref{lemma:independence-preserve}. The inequality above,
 along with the binding condition \eqref{e:BC_secrecy2},
 yields
 \begin{align*}
 1-\epsilon &\leq \bPr{\hat{K}=K} + \bPr{T(\hat{K},
   \hat{X}_1, X_2,\bF) = 0, \hat{K}\neq K} \\ &\leq
 \bPr{\hat{K}=K} + \delta_2,
 \end{align*}
 which completes the proof of \eqref{e:BC_to_show}.  \qed

We conclude this section by observing a simple application
of Theorem \ref{theorem:BC_single_shot_bound} in bounding the
efficiency of reduction of BC to OT. For a detailed
discussion, see \cite{RanTapWinWul11}.

\begin{example}[{{\bf Reduction of BC to OT}}] Suppose two parties have at their disposal an OT of length $n$. Using this as a resource, what is the length $l$ of $(\ep, \delta_1, \delta_2)$-BC that can be constructed? 

Denoting by $K_0, K_1$ the OT strings, and by $B$ the OT bit
of Party 2, let $X_1= (K_0, K_1)$ and $X_2=(B, K_B)$. Note
that \eqref{eq:no-dispersion-condition} holds with
$\mathrm{P}= \bPP{X_1X_1X_2}$ and $\mathrm{Q} = \bPP{X_1 |
  X_2}\bPP{X_1X_2}$, and
$$D(\bPP{X_1X_1X_2} \| \bPP{X_1 | X_2}\bPP{X_1X_2}) = n.$$
Therefore, by Theorem \ref{theorem:BC_single_shot_bound} and
\eqref{eq:no-dispersion-bound}, we get
$$l \leq n + \log(1/(1-\epsilon-\delta_1-\delta_2-\eta)) + 2
\log (1/\eta),$$ where $0< \eta < 1-
\epsilon-\delta_1-\delta_2$. This bound on efficiency of
reduction is stronger than the one derived in
\cite[Corollary 2]{RanTapWinWul11} (fixing $n=n^\prime = 1$
in that bound). In particular, it shows an additive loss of
logarithmic order in $(1-\ep-\delta_1-\delta_2)$, while
\cite[Corollary 2]{RanTapWinWul11} shows a multiplicative
loss of linear order.
\end{example}

\section{Implications for secure computation with trusted parties}\label{section:secure-computing-trusted}

In this section, we present a connection of our result to a
problem of secure function computation with trusted parties,
where the parties seek to compute a function of their
observations using a communication that does not reveal the
value of the function by itself (without the observations at
the terminals). This is in contrast to the secure
computation treated in Section
\ref{section:secure-computing} where the communication is
secure but the parties are required not to get any more
information than the computed function value. This problem
was introduced in \cite{TyaNarGup11} where a matching
necessary and sufficient condition was given for the
feasibility of secure computation in the asymptotic case
with IID observations. Here, using Theorem
\ref{theorem:one-shot-converse-source-model}, we derive a
necessary condition for the feasibility of such secure
computing for general observations (not necessarily IID).

Formally, consider $m\geq 2$ parties observing RVs $X_1,
..., X_m$ taking values in finite sets $\cX_1, ..., \cX_m$,
respectively. Upon making these observations, the parties
communicate interactively in order to {\it securely compute}
a function $g: \cX_1\times...\times \cX_m \rightarrow \cG$
in the following sense: The $i$th party forms an estimate
$G_{(i)}$ of the function based on its observation $X_i$,
local randomization $U_i$ and interactive communication
$\bF$, $i.e.$, $G_{(i)} = G_{(i)}(U_i, X_i, \bF)$. For $0
\leq \ep,\delta < 1$, a function $g$ is {\it $(\ep,
  \delta)$-securely computable} if there exists a protocol
satisfying
\begin{align}
\bPr{G = G_{(1)} = \cdots = G_{(m)}} &\geq 1- \ep,
\label{e:secure-computability-1}
\\ \d{\bPP{G\bF}}{\bPP{G}\times \bPP{\bF }} &\leq \delta,
\label{e:secure-computability-2}
\end{align}
where $G = g\left(X_\cM\right)$. The first condition
captures the reliability of computation and the second
condition ensures the secrecy of the
protocol. Heuristically, for secrecy we require that an
observer of (only) $\bF$ must not get to know the computed
value of the function.  We seek to characterize the $(\ep,
\delta)$-securely computable functions $g$.

In \cite{TyaNarGup11}, an asymptotic version of this problem
was addressed. The parties observe $X_1^n, ..., X_m^n$ and
seek to compute $G_t = g\left(X_{1t}, ..., X_{mt}\right)$
for each $t\in \{1, ..., n\}$; consequently, the RVs $\{G_t,
1 \leq t \leq n\}$ are IID. A function $g$ is securely
computable if the parties can form estimates $G_{(1)}^{(n)},
..., G_{(m)}^{(n)}$ such that
\begin{align*}
\bPr{G^n = G_{(1)}^{(n)} = \cdots = G_{(m)}^{(n)}} &\geq 1 -
\ep_n, \\
\d{ \bPP{G^n\bF}}{ \bPP{G^n}\times \bPP{\bF}}
&\leq \ep_n,
\end{align*}
where $\displaystyle\lim_{n\rightarrow \infty} \ep_n =
0$. The following characterization of securely computable
functions $g$ is known.
\begin{theorem}\cite{TyaNarGup11} For the asymptotic case described above, a function $g$ is securely computable if 
$H(G) < C$, where $H(G)$ is the entropy of the RV $G =
  g(X_\cM)$ and $C = C(X_\cM)$ is the SK capacity given in
  Theorem \ref{theorem-capacity}.

Conversely, if a function $g$ is securely computable, then
$H(G) \leq C$.
\end{theorem}
Heuristically, the necessary condition above follows upon
observing that if the parties can securely compute the
function $g$, then they can extract a SK of rate $H(G)$ from
RVs $G^n$. Therefore, $H(G)$ must be necessarily less than
the maximum rate of a SK that can be generated, namely the
SK capacity $C$.  Using this heuristic, we present a
necessary condition for a function $g$ to be $(\ep,
\delta)$-securely computable.

\begin{corollary}\label{corollary:secure-computability-necessary-single-shot}
For $0 \leq \ep, \delta <1$ with $\ep+\delta <1$, if a
function $g$ is $(\ep, \delta)$-securely computable, then
\begin{align}
H^{\xi}_{\min}(\bPP{G}) \leq \frac{1}{|\pi|-1} \bigg[-\log
  \beta_{\mu}\big(\bPP{X_{\cM}},\mathrm{Q}^\pi_{X_{\cM}}\big)
  + |\pi| \log(1/\eta)\bigg]+ 2\log(1/2\zeta) + 1,
~\forall
\,\mathrm{Q}^\pi_{X_{\cM}} \in {\cal Q}(\pi),
\label{e:secure-computability-necessary-single-shot}
\end{align}
for every $\mu = \ep+\delta+2\xi + \zeta+\eta$ with $\xi,
\zeta, \eta >0$ such that $\mu < 1$, and for every partition
$\pi$ of $\cM$.
\end{corollary}
{\it Proof.} The proof is based on extracting an $\ep$-SK
from the RV $G$ that the parties share.  Specifically, Lemma
\ref{lemma:left-over-hash} with $X = G$, $Y = \mbox{const}$,
and condition (\ref{e:secure-computability-2}) imply that
there exists $K = K(G)$ with $\log|\cK| = \lfloor
H^{\xi}_{\min}(\bPP{G}) - 2\log(1/2\zeta)\rfloor$ and
satisfying
\begin{eqnarray*}
\lefteqn{
  \d{\mathrm{P}_{K(G)\bF}}{\mathrm{P}_{\mathrm{unif}} \times
    \mathrm{P}_{\bF}} } \\ &\le&
\d{\mathrm{P}_{K(G)\bF}}{\mathrm{P}_{K(G)} \times
  \mathrm{P}_{\bF}} + \d{\mathrm{P}_{K(G)} \times
  \mathrm{P}_{\bF}}{\mathrm{P}_{\mathrm{unif}} \times
  \mathrm{P}_{\bF}} \\ &\le& \d{
  \mathrm{P}_{G\bF}}{\mathrm{P}_{G} \times \mathrm{P}_{\bF}}
+ \d{\mathrm{P}_{K(G)}}{\mathrm{P}_{\mathrm{unif}}} \\ &\le&
\delta + 2\xi + \zeta.
\end{eqnarray*}
Thus, in the view of Proposition
\ref{proposition:sec_equiv}, the RV $K$
constitutes\footnote{Strictly speaking, the estimates $K_1,
  ..., K_m$ of $K$ formed by different parties constitute
  the $(\ep + \delta + 2\xi+\zeta)$-SK in the sense of
  (\ref{e:secrecy}).} an $(\ep + \delta + 2\xi +\zeta)$-SK.
An application of Theorem
\ref{theorem:one-shot-converse-source-model} gives
(\ref{e:secure-computability-necessary-single-shot}).\qed

We conclude this section with two illustrative examples.
\begin{example}{\bf (Computing functions of independent observations using a perfect SK).}
Suppose the $i$th party observes $U_i$, where the RVs $U_1,
..., U_m$ are mutually independent. Furthermore, all parties
share a $\kappa$-bit perfect SK $K$ which is independent of
$U_\cM$. How many bits $\kappa$ are required to $(\ep,
\delta)$-securely compute a function $g\left(U_1, ...,
U_m\right)$?

Note that the data observed by the $i$th party is given by
$X_i= \left(U_i, K\right)$. A simple calculation shows that
for every partition $\pi$ of $\cM$,
\begin{eqnarray*} 
\beta_\ep\left(\bPP{X_{\cM}},\prod_{i=1}^{|\pi|}
\bPP{X_{\pi_i}}\right) \geq (1-\ep)\kappa^{1- |\pi|},
\end{eqnarray*}
and therefore, by Corollary
\ref{corollary:secure-computability-necessary-single-shot} a
necessary condition for $g$ to be $(\ep, \delta)$-securely
computable is
\begin{align}
H^{\xi}_{\min}(\bPP{G}) \leq \kappa +
\frac{1}{|\pi|-1}\left(|\pi|\log(1/\eta) +
\log(1/(1-\mu))\right) + 2\log(1/2\zeta) + 1,
\label{e:necessary-condition-secure-computing-shared-key}
\end{align}
for every $\xi,\zeta,\eta>0$ satisfying $\mu =\ep + \delta +
2\xi + \zeta + \eta <1$. Note that the finest partition,
i.e., $|\pi| = m$, gives the tightest bound in
\eqref{e:necessary-condition-secure-computing-shared-key}.

For the special case when $U_i = B_i^n$, a sequence of
independent, unbiased bits, and
$$g\left(B_1^n, ..., B_m^n\right) = B_{11}\oplus...\oplus
B_{m1}, ..., B_{1n}\oplus...\oplus B_{mn},$$ $i.e.$, the
parties seek to compute the (element-wise) parities of the
bit sequences, it holds that $H^{\xi}_{\min}(\bPP{G}) \geq
n$.  Therefore, $(\ep, \delta)$-secure computation is
feasible only if $n \leq \kappa + O(1)$.  We remark that
this necessary condition is also (almost)
sufficient. Indeed, if $n \leq \kappa$, all but the $m$th
party can reveal all their bits $B_1^n,\ldots,B_{m-1}^n$ and
the $m$th party can send back $B_1^n\oplus \ldots\oplus
B_m^n\oplus K_n$, where $K_n$ denotes any $n$ out of
$\kappa$ bits of $K$. Clearly, this results in a secure
computation of $g$.
\end{example}

\begin{example}{\bf(Secure transmission).}
Two parties sharing a $\kappa$-bit perfect SK $K$ seek to
exchange a message $M$ securely.\footnote{A message $M$ is a
  RV with known distribution $\bPP{M}$.} To this end, they
communicate interactively using a communication $\bF$, and
based on this communication Party 2 forms an estimate
$\hat{M}$ of the message $M$ by Party 1. This protocol
accomplishes $(\ep, \delta)$-secure transmission if
\begin{align*}
\bPr{M = \hat{M}} &\geq 1- \ep, \\
\d{\bPP{M\bF}}{\bPP{M}\times \bPP{\bF }} &\leq \delta.
\nonumber
\end{align*}
The classic result of Shannon \cite{Sha49} implies that
$(0,0)$-secure transmission is feasible only if $\kappa$ is
at least $\log\|M\|$, where $\|M\|$ denotes the size of the
message space.\footnote{This is a slight generalization of
  Shannon's original result; see \cite[Theorem
    2.7]{KatLin07} for a proof.} But, can we relax this
constraint for $\ep, \delta >0$? In this example, we will
give a necessary condition for the feasibility of $(\ep,
\delta)$-secure transmission by relating it to the previous
example.

Specifically, let the observations of the two parties
consist of $X_1 = (M, K)$, $X_2 = K$. Then, $(\ep,
\delta)$-secure transmission of $M$ is tantamount to
securely computing the function $g(X_1, X_2) =
M$. Therefore, using
(\ref{e:necessary-condition-secure-computing-shared-key}),
$(\ep, \delta)$-secure transmission of $M$ is feasible only
if
\begin{align}
H^{\xi}_{\min}(\bPP{M}) \leq \kappa + 2\log(1/\eta) +
\log(1/(1-\mu)) + 2\log(1/2\zeta) + 1,
\label{e:necessary-condition-for-one-time-pad-general}
\end{align}
for every $\xi,\zeta,\eta>0$ satisfying $\mu =\ep + \delta +
2\xi + \zeta + \eta <1$.

Condition
(\ref{e:necessary-condition-for-one-time-pad-general})
brings out a trade-off between $\kappa$ and $\ep + \delta$
(cf.~\cite[Problems 2.12 and 2.13]{KatLin07}). For an
illustration, consider a message $M$ consisting of a RV $Y$
taking values in a set $\cY = \{0,1\}^n \cup \{0,1\}^{2n}$
and with the following distribution:
\begin{eqnarray*}
\bPP{Y}(y) = \left\{
\begin{array}{ll}
\frac{1}{2} \cdot \frac{1}{2^n} & y \in \{0,1\}^n
\\ \frac{1}{2} \cdot \frac{1}{2^{2n}} & y \in \{0,1\}^{2n}
\end{array}
\right..
\end{eqnarray*}
For $\ep + \delta =0$, we know that secure transmission will
require $\kappa$ to be more than the {\it worst-case message
  length} $2n$. But perhaps by allowing $\ep + \delta$ to be
greater than $0$, we can make do with fewer SK bits; for
instance, perhaps $\kappa$ equal to $H(M) = (3/2)n +1$ will
suffice (note that the {\it average message length} equals
$(3/2)n$). The necessary condition above says that this is
not possible if $\ep+\delta < 1/2$. Indeed, since
$H_{\min}^\xi(\bPP{Y}) \ge 2n$ for $\xi = 1/4$, we get from
(\ref{e:necessary-condition-for-one-time-pad-general}) that
the message $M=Y$ can be $(\ep, \delta)$-securely
transmitted only if $2n \le \kappa + O(1)$, where the
constant depends on $\ep$ and $\delta$.
\end{example}

\section{Discussion}

In this work, we focused on converse results and presented
single-shot upper bounds on the efficiency of using
correlated randomness for SK agreement and secure
computation protocols. When the underlying observations are
IID, the resulting upper bounds were shown to be tight in
several cases. It is natural to ask how tight are these
bounds for IID observations of fixed, finite length. For the
SK agreement problem, it is possible to mimic the approach
in \cite{Mau93,AhlCsi93,CsiNar04,RenWol05} to obtain
protocols that first use communication for {\it information
  reconciliation} and then extract SKs using {\it privacy
  amplification}.  The challenge in the multiparty setup is
to identify the appropriate {\it information to be
  reconciled}.  For the case of two parties observing IID
sequences, relying on Theorem
\ref{theorem:one-shot-converse-source-model}, recently the
second-order asymptotic term in the maximum length of a SK was
established in \cite{HayTyaWat14ii, HayTyaWat14}. Coming up
with finite-length schemes that match the converse bounds
for the various secure computation problems studied above is
work in progress.

Our converse results in Sections
\ref{section:secure-computing} and
\ref{section:secure-computing-trusted} entail reducing SK
agreement to the secure computation task at hand, followed
by an application of Theorem
\ref{theorem:one-shot-converse-source-model}.  The strength
of Theorem~\ref{theorem:one-shot-converse-source-model} lies
in its validity for interactive communication. The
admissibility of interactive communication makes this bound
useful in cryptography where interaction is natural to
consider, and it is foreseeable, and indeed tempting, that
this approach can lead to converse bounds for other problems
in information theoretic secrecy and cryptography; an instance arises in
\cite{HayTyaWat14iii}.

In fact, our bound can find applications in problems
involving interactive communication without any secrecy
requirements.  For instance, it is used in
\cite{TyaVisWat15} to derive a lower bound for the length of
the interactive communication needed for two parties to exchange
their correlated data. Furthermore, it is used in
\cite{TyaVenVisWat15} to derive a lower bound on
the communication complexity for simulating protocols.

Note that similar to \cite{PolPooVer10, HayNag03} 
the choice of $\bQQ{}$ in 
Theorem \ref{theorem:one-shot-converse-source-model} 
is arbitrary. In the applications to capacity results 
considered in 
this paper and in deriving the second-order asymptotics
for the two party SK agreement problem
in~\cite{HayTyaWat14ii}, 
$\bQQ{}$ equal
to the product of marginals of $\bPP{}$ suffices. 
However,
in a more involved application of Theorem
\ref{theorem:one-shot-converse-source-model},
such as that in~\cite{TyaVisWat15,TyaVenVisWat15},
a judicious choice of $\bQQ{}$ is needed.

A quantum version of the
two party secret key agreement problem of
\cite{Mau93, AhlCsi93} has been considered in
\cite{DevWin05, CEHHOR07}. An extension of
Theorem \ref{theorem:one-shot-converse-source-model}
to the case of quantum observations can be used to
obtain converse results for such problems.
In the classical case, for two parties with IID
observations,
Theorem~\ref{theorem:one-shot-converse-source-model} 
shows that the $(\ep,\delta)$-SK capacity is bounded above by
\begin{align} \label{eq:identity}
\min_{\bQQ{X_1|Z}\bQQ{X_2|Z}\bQQ{Z}} D(\bPP{X_1 X_2 Z} \|
\bQQ{X_1|X}\bQQ{X_2|Z}\bQQ{Z}) = I(X_1 \wedge X_2 | Z),
\end{align}
where the equality follows from the Tops\o e
identity \cite{Top67}. On the other hand, in the quantum case, 
the identity \eqref{eq:identity} does not
hold \cite{IbiLinWin08}. Thus, a direct extension of
Theorem~\ref{theorem:one-shot-converse-source-model} to 
quantum observations will not yield
the quantum conditional mutual information bound 
for SK capacity derived in \cite{CEHHOR07}.
Finding an appropriate extension of
Theorem~\ref{theorem:one-shot-converse-source-model} 
to the case of quantum observations is an interesting
direction
for future research.


%

\bibliography{IEEEabrv,references}
\bibliographystyle{IEEEtranS}

\end{document}